\documentclass[sigconf,screen]{acmart}
\AtBeginDocument{%
  \providecommand\BibTeX{{%
    \normalfont B\kern-0.5em{\scshape i\kern-0.25em b}\kern-0.8em\TeX}}}

\usepackage{listings}
\usepackage{xcolor}
\usepackage{subcaption}

\lstdefinestyle{haskell}{
  frame=none,
  xleftmargin=2pt,
  stepnumber=1,
  numbers=left,
  numbersep=5pt,
  numberstyle=\ttfamily\tiny\color[gray]{0.3},
  belowcaptionskip=\bigskipamount,
  alsoletter={=<>-|{}},
  captionpos=b,
  escapeinside={*'}{'*},
  morekeywords={do, if, then, else, case, of, class, data, newtype, instance, where, deriving, import, let, in, module, qualified, type},
  otherkeywords={<-, ->, ::, \$, [, ], \,, =, |, \{, \}, (, )},
  tabsize=2,
  emphstyle={\bf},
  comment=[l]{--},
  commentstyle=\color{red} \it,
  stringstyle=\mdseries\rmfamily,
  showspaces=false,
  columns=flexible,
  basicstyle=\small\sffamily,
  showstringspaces=false,
  morecomment=[l]\%,
}
\newcommand{\code}[1]{\texttt{#1}}

\copyrightyear{2023}
\acmYear{2023}
\setcopyright{rightsretained}
\acmConference[IFL 2023]{The 35th Symposium on Implementation and Application of Functional Languages}{August 29--31, 2023}{Braga, Portugal}
\acmBooktitle{The 35th Symposium on Implementation and Application of Functional Languages (IFL 2023), August 29--31, 2023, Braga, Portugal}\acmDOI{10.1145/3652561.3652570}
\acmISBN{979-8-4007-1631-7/23/08}

\begin{document}

\title{QuickerCheck}
\subtitle{Implementing and Evaluating a Parallel Run-Time for QuickCheck}

\author{Robert Krook}
\email{krookr@chalmers.se}
\affiliation{%
  \institution{Chalmers University of Technology}
  \city{Gothenburg}
  \country{Sweden}
}

\author{Nicholas Smallbone}
\email{nicsma@chalmers.se}
\affiliation{%
  \institution{Chalmers University of Technology}
  \city{Gothenburg}
  \country{Sweden}
}

\author{Bo Joel Svensson}
\email{bo.joel.svensson@gmail.com}
\affiliation{
  \institution{Lind Art \& Technology}
  \city{Stockholm}
  \country{Sweden}
}

\author{Koen Claessen}
\email{koen@chalmers.se}
\affiliation{%
  \institution{Chalmers University of Technology}
  \city{Gothenburg}
  \country{Sweden}
}

\begin{abstract}
This paper introduces a new parallel run-time for \emph{QuickCheck}, a Haskell library and EDSL for specifying and randomly testing properties of programs. The new run-time can run multiple \emph{tests} for a single property in parallel, using the available cores. Moreover, if a counterexample is found, the run-time can also \emph{shrink} the test case in parallel, implementing a parallel search for a locally minimal counterexample.

Our experimental results show a 3--9$\times$ speed-up for testing QuickCheck properties on a variety of heavy-weight benchmark problems. We also evaluate two different shrinking strategies; \emph{deterministic shrinking}, which guarantees to produce the same minimal test case as standard sequential shrinking, and \emph{greedy shrinking}, which does not have this guarantee but still produces a locally minimal test case, and is faster in practice.
\end{abstract}

\ccsdesc[500]{Computing methodologies~Concurrent algorithms}
\ccsdesc[500]{Computing methodologies~Shared memory algorithms}
\ccsdesc[500]{Theory of computation~Shared memory algorithms}
\ccsdesc[500]{Software and its engineering~Software testing and debugging}

\keywords{property-based testing, quickcheck, testing, parallel functional programming, haskell}

\maketitle

\section{Introduction}
\label{sec:introduction}

QuickCheck \cite{DBLP:conf/icfp/ClaessenH00} is a widely known Haskell tool for property-based random testing of programs. First, the programmer writes a property of the program under test that they expect to always hold. Then, to check the property, QuickCheck generates a number of random test cases to exercise the property. If the property always held, the check is reported as successful. If a test case makes the property fail, a process called \emph{shrinking} is invoked, which consists of a greedy search for a (locally) minimal failing test case.

For example, we may be testing an implementation of System F \cite{girard1972interpretation}, where we have the following types and functions:
\begin{lstlisting}
type Expr -- expressions
type Type -- types

reduce :: Expr -> Maybe Expr
typeOf :: Expr -> Type
\end{lstlisting}
The types \code{Expr} and \code{Type} stand for expressions and types of expressions in System F. The function \code{reduce} takes one evaluation step, if possible. The function \code{typeOf} computes the type of an expression. \emph{Subject reduction} is a property that says that evaluation of expressions does not cause their type to change. This can be expressed as a QuickCheck property as follows:
\begin{lstlisting}
prop_Preservation :: Expr -> Property
prop_Preservation e =
  isJust r ==> typeOf e == typeOf (fromJust r)
  where
    r = reduce e
\end{lstlisting}
Here, the operator \code{==>} specifies a \emph{precondition}: only tests satisfying \code{isJust r} are of interest.

To run QuickCheck, the user must also supply an \code{Arbitrary} instance
describing how to generate random \emph{well-typed} \code{Expr}s\footnote{There is no requirement by QuickCheck itself that the generator has to generate well-formed terms. This is primarily required to meaningfully exercise the property in question.} (a non-trivial
task studied in e.g. \cite{palka2011testing}). QuickCheck will then generate a configurable amount of random expressions, which by default is 100, and evaluate the property for them. In fact, QuickCheck will
typically evaluate the property even more times, because:
\begin{itemize}
\item QuickCheck discards any test case not satisfying the precondition
  \code{isJust r}, and continues until it has executed 100 tests satisfying the precondition.
\item If a test case fails (for example if the function \code{reduce} contains
  a bug), shrinking searches for a smaller counterexample by executing the property on many smaller test cases.
\end{itemize}

All this happens \emph{sequentially} at the moment in QuickCheck. If the evaluation of the property takes a long time, QuickChecking it (and possibly shrinking the counterexample) will take an even longer time. This is time often spent waiting by the programmer, perhaps wondering why their computer is roaring like a spaceship while only one core is in use. The contribution of this paper is to propose and practically evaluate a way of performing both the testing phase as well as the shrinking phase of QuickCheck in parallel\footnote{The implementation can be found at \href{https://github.com/Rewbert/quickcheck}{https://github.com/Rewbert/quickcheck}. The authors intend to eventually merge this work into mainline QuickCheck.}.

Note that our work aims to reduce waiting time for the programmer while checking a \emph{single property}. There exist frameworks (for example tasty \cite{tasty}) that allow testing of multiple properties and unit tests in parallel or even distributed on a cluster. These are typically used in regression tests or continuous integration. 
Our work can not only speed up testing in these settings but also during active development, where programmers typically run QuickCheck on a single property and wait for the result.

\section{What are the challenges?}

Even though running each test is supposed to be independent, and as such testing a property 100 times should be a so-called \emph{embarrassingly parallel} task, in practice parallelizing testing is not so easy. For one, individual tests may interact with each other, but luckily in Haskell, we often get the independence guarantees we need from pure (or at least thread-safe) code. In this paper, we assume that the property itself is thread-safe.

But the biggest problem is that \emph{QuickCheck's algorithm is inherently sequential}. This is not at all obvious at first glance. The problem comes from
two features in QuickCheck -- adjustment of test size, and shrinking. As we will see, these features introduce a data dependency: the test case that we should try next depends on the result of the \emph{previous} test. Addressing these dependencies was one of the main challenges in parallelizing QuickCheck.

\paragraph{Test size}

Many times it is enough to generate a small test input to falsify a property. QuickCheck tries to generate smaller inputs early on, and gradually increases the size as more and more tests are passed. This is achieved by QuickCheck supplying the test-data generator with a \textit{size}. The generators are free to disregard the size completely but may use it if they wish to. As an example, the default generator for lists uses the size as an upper bound of the length of the generated list. The size of the first test is always 0, while the default upper bound is 100. If the user specifies that 100 tests should be executed, QuickCheck will make sure that the generator has been provided with all sizes between 0 and 99. The authors point out that so far everything discussed is easily parallelisable.

However, properties in QuickCheck can not only succeed or fail, but also \emph{discard}, which means that a pre-condition in the property was not fulfilled. A discarded test case is not counted towards the total number of successful tests\footnote{The reason for this is that if the precondition of a property is more likely to succeed for small test data sizes, we still want to make sure that we exercise the property on larger sizes.}. So, the distribution of sizes during testing only depends on the total number of \emph{successful} tests so far, not on the total number of tests in general. This introduces a small but significant data dependency preventing parallel evaluations of tests; when a worker runs a test it must know the appropriate size of the test case to run, and the appropriate size depends on whether the previous tests were successful or not. This dependency needs to be dealt with somehow in the parallelization.

\paragraph{Shrinking}

\sloppy
When a failing test case is found, QuickCheck searches for a smaller failing
test case by applying a process called \emph{shrinking} \cite{DBLP:journals/tse/ZellerH02}. The goal
of shrinking is to produce a locally minimal failing test case. Shrinking first
produces a list of \emph{shrink candidates}, variants of the test case that
have been reduced in size in a variety of ways. This list is traversed from left
to right until we find a new failing test case. Shrinking is then applied
recursively on the new failing test case until the current failing test case
cannot be reduced anymore. Shrinking does not backtrack in search of a globally minimal counterexample, but only promises to yield a local minimum.

\fussy
To use shrinking, the user must define a \emph{shrink function}. For a type \code{T}, this is a function
\code{shrink :: T -> [T]} which, given a test case, produces a list of shrink candidates, i.e. smaller or simpler test cases to try. For example, suppose that we are testing System F again, and the type of expressions is defined as follows:

\begin{lstlisting}
data Expr
  = Var String    -- variable
  | App Expr Expr -- application
  | ...           -- other constructors
\end{lstlisting}

Then we can define a shrink function as follows:

\begin{lstlisting}
shrink :: Expr -> [Expr]
shrink (Var _) = []
shrink (App t u) =
  concat [ [ t, u ]
         , [ App t' u | t' <- shrink t ]
         , [ App t u' | u' <- shrink u ]
         ]
shrink (...) = ... -- other constructors
\end{lstlisting}
A \code{Var} can not be shrunk further, so we return an empty list of candidates. A function application \code{App t u}, however, can be shrunk further.
We can remove the \code{App} constructor and one of the subexpressions, leaving
just \code{t} or \code{u}, which if successful may shrink the expression
considerably. We can also keep the \code{App} constructor but shrink the
subexpressions.

Note that QuickCheck \emph{always tries the shrink candidates in the order they appear in the list, from left to right}. Hence it is common to return the \textit{greedy} candidates first, those that remove large parts of the value,
as we do in this case. Ordering the shrink list appropriately can greatly improve the speed of shrinking.

We propose to parallelize shrinking in two ways:
\begin{enumerate}
\item \emph{Greedy shrinking} evaluates as many shrinking candidates in parallel as possible, and as soon as a candidate fails, it recursively continues with that candidate. It may be that a candidate \emph{earlier} in the shrink list (corresponding to a more aggressive shrink step) would also have failed if we had waited, and in that case, we may perform a smaller shrink step than necessary.
\item \emph{Deterministic shrinking} speculatively evaluates test cases in the
search before we know we will need to, but always makes the same choices as in
the sequential case. That is, when a shrink candidate fails, it waits until it knows that \emph{no earlier candidate fails}
\end{enumerate}
In our evaluation, greedy shrinking is usually faster than deterministic shrinking.

\section{QuickerCheck}

We present QuickerCheck via two examples. We point out that the QuickCheck API for writing generators, shrinkers, and properties remains unchanged, and only the internal evaluation of a property is modified.

\paragraph{System F}
In Section~\ref{sec:introduction}, we saw the property
\code{prop\_Preservation :: Expr -> Property} for testing subject reduction in System F. To test this property with \emph{sequential} QuickCheck we run:

\begin{verbatim}
> quickCheck prop_Preservation
+++ OK! Passed 100 tests.
\end{verbatim}

As the property is pure, it is safe to test in parallel using QuickerCheck.
To do so, we must compile the code with the \code{-threaded} and \code{-rtsopts} flags and pass in the \code{-N} option to the run-time system, to enable parallelism in GHC.
Then all we have to do is invoke \code{quickCheckPar} instead of \code{quickCheck}.

The output (assuming all tests passed) is
\begin{verbatim}
> quickCheckPar prop_Preservation
+++ OK! Passed 100 tests.
  tester 0: 50
  tester 1: 50
\end{verbatim}

The lines \texttt{tester 0: 50} and \texttt{tester 1: 50} show that two threads were used (we happened to limit GHC to using two cores) and that they each executed 50 test cases. What is not visible in the output is that, since the tests were distributed among two cores, QuickerCheck ran close to twice as fast.

\paragraph{Compiler testing}
\label{para:compilertesting}
A function that is not necessarily embarrassingly parallel is one that is effectful. To test a compiler it is necessary to perform IO actions, such as invoking the compiler under test or executing the compiled binary. Testing compilers is non-trivial, but a well-studied approach is \emph{metamorphic testing} \cite{DBLP:journals/corr/abs-2002-12543}. In this approach, assuming a function of type \code{Program -> IO Output} that compiles and runs the program, we define a function \code{mutateProgram :: Program -> Program} that mutates the program in some way, and then specify how the output should change in response by a function \code{mutateOutput :: Output -> Output}. Mathematically, the property that should hold is:

\begin{lstlisting}
-- compileAndRun :: Program -> IO Output
compileAndRun (mutateProgram p) =
  fmap mutateOutput (compileAndRun p)
\end{lstlisting}

In practice, we also need to perform various housekeeping tasks such as writing
the program source to a file and cleaning up output files, so a more realistic
property is:

\begin{lstlisting}
prop_metamorphic :: Program -> Property
prop_metamorphic program = ioProperty $ do
  writeFile "p.c" (render program)
  writeFile "q.c" (render $ mutateInput program)
  output1 <- compileAndRun "p.c"
  output2 <- compileAndRun "q.c"
  mapM removeFile ["p.c", "q.c", "p.exe", "q.exe"]
  return (mutateOutput output1 == output2)
\end{lstlisting}

The property executes both the original and modified programs after having first written them to the file system. The file system is cleaned up, after which the outputs are compared. The output of the unmodified program is modified to reflect the change described by the metamorphic relation.

Unfortunately, running this property with \code{quickCheckPar} will produce \emph{extremely strange} test failures. The reason is that the property, while innocent-looking, is not thread-safe. There is an implicitly shared resource, the file system: if multiple instances of the property execute in parallel, they will all write to the \emph{same} files \texttt{p.c} and \texttt{q.c}. This leads to obvious race conditions. There are different ways to modify the property such that there are no race conditions, one of which is to let the property create a temporary directory to which intermediary files are written.

\begin{lstlisting}
-- create a fresh temporary directory based on a baseline name
-- withSystemTempDirectory :: String -> (FilePath -> IO a) -> IO a

prop_metamorphic :: Program -> Property
prop_metamorphic program = ioProperty $ do
  withSystemTempDirectory "compiler_output" $ \dir -> do
    -- rest of property, now using dir as a
    -- scratch space for temporary files
\end{lstlisting}

If we disregard other implicitly shared resources such as CPU caches, RAM, bandwidth, etc, this property can now be evaluated in parallel by using \code{quickCheckPar}.

In general, using QuickerCheck requires three steps. (1) Make sure that the property is thread-safe (only for properties doing I/O). (2) Compile the program with threading options. (3) Run \code{quickCheckPar} instead of \code{quickCheck}.

\section{QuickerCheck Design and Implementation}
\label{sec:design}

The extensions to QuickCheck described in this paper are designed such that as few observable behaviors as possible are changed. Some design choices of QuickCheck do not lend themselves nicely to parallelism, and QuickerCheck tries to make reasonable compromises where possible. One notable case of this is the way QuickCheck computes sizes for a test case. The size is derived from the number of tests that have passed so far, and the number of tests that have been discarded since the last passing test. This means that we can not compute the size of a test until we have observed the outcome of all tests that came before. This sounds sub-optimal for parallelization; below, we explain what QuickerCheck does to address this.

\paragraph{Testing}

The test loop in ordinary, sequential QuickCheck is a recursive function that
maintains a state containing e.g. the count of how many tests were executed so
far, how many were discarded due to a failed pre-condition, etc. It also holds
the random seed used to generate the test case. It generates and executes one
test at a time, adjusting the size of the test case whenever a test succeeds,
but not when a test is discarded. Once a test fails, the test loop terminates and a shrinking routine is invoked.

The parallel test loop is implemented by running concurrent instances of the sequential test loop. The main thread spawns concurrent \textit{testers} that evaluate one test after another, and then goes to sleep until the testers report that all tests have been executed, too many tests were discarded, or a counterexample was found. The test loop maintains a state that is updated after every test, recording how many tests have been passed so far, the next random seed, and many other things. In order to facilitate multiple, concurrent testers, some of the state has been moved into \code{MVar}s. As an example, the integer representing the number of tests a particular thread has yet to run resides in an \code{MVar}, enabling other threads to read it if they wish to steal work from that thread.

\begin{figure}
        \includegraphics[width=0.22\textwidth]{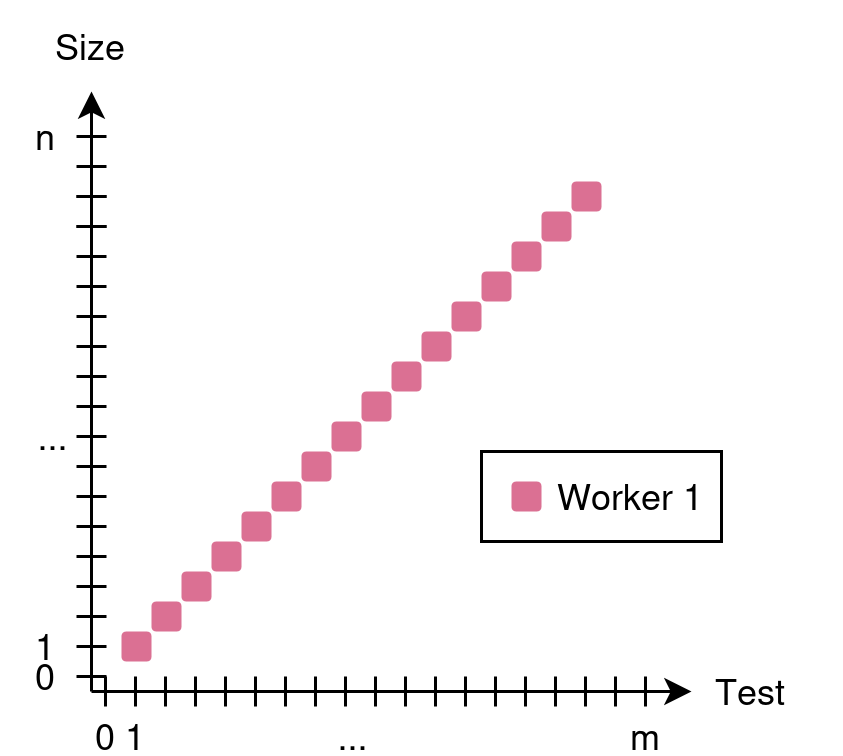}
        \includegraphics[width=0.22\textwidth]{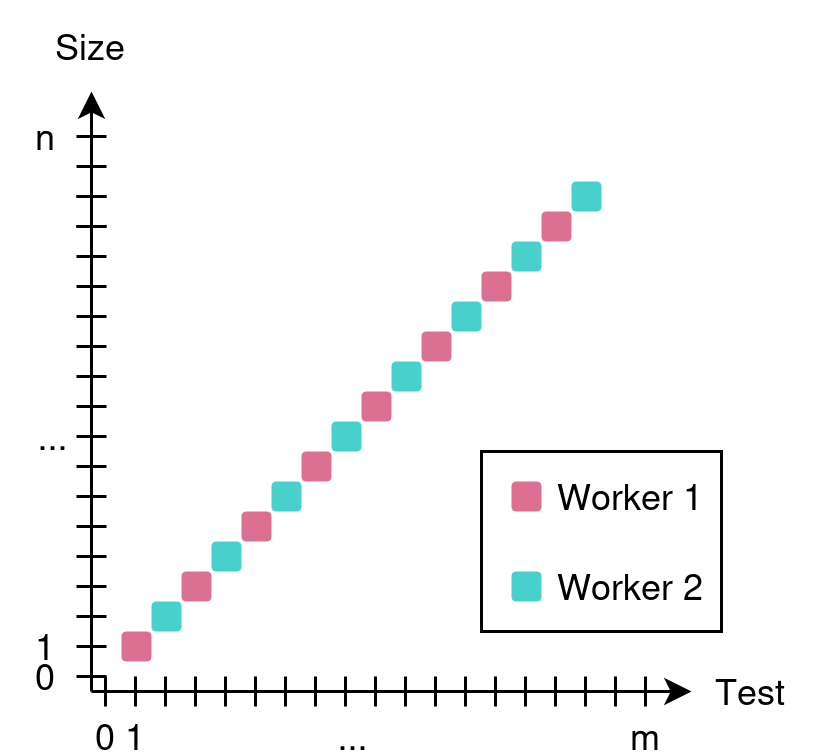}

    \caption{An illustration of how size grows as more and more tests are passed, and to which worker they are assigned. In order to get a fair workload for the concurrent testers a stride is applied when computing sizes.}
    \label{fig:sizecomputation}
\end{figure}

Communication between threads occurs as little as possible in order to not incur synchronization costs. When testing is initiated, the number of tests to run is divided equally between the testers, and only when one thread has exhausted its budget of tests will it inspect the budgets of the concurrent testers. If work-stealing is enabled, a thread may then decrement the counter of a sibling tester and run another test on its own. Each tester has its own random seed that it splits before running a test, as sharing a seed between all testers would incur synchronization overheads. Additionally, each tester computes the sizes to use for test cases based on their individual counters for how many tests they have passed, and how many tests they have discarded since the last passing test. In an effort to explore the same set of sizes as in sequential QuickCheck,
they each apply a \emph{stride}: If we have $k$ threads, then thread number $i$ uses sizes \(i, i+k, i+2k, i+3k, \dots\). This is illustrated in figure \ref{fig:sizecomputation}. A compromise is made when a thread steals a test from a sibling tester, in which case the local next size is used. This reduces synchronization costs, as the thread that ran the test doesn't need to report the result back to the other thread. With this approach, we explore the same set of sizes as sequential QuickCheck, except when work stealing happens.

As an alternative to strides, we have also implemented a strategy that divides the set of sizes into contiguous segments for each of the testers, by applying an offset to the size computation. There is a risk, however, that test cases generated by e.g. smaller sizes will run faster than test cases generated with larger sizes. This would lead to the concurrent testers finishing their given workloads at different times. Computing sizes with an offset is implemented and can be chosen by configuring the arguments to \code{quickCheckWith}\footnote{The function \code{quickCheckWith} is a variant of \code{quickCheck} that accepts a configuration parameter where default behavior can be overridden.}, but the default behavior is to use a stride.

When a thread finds a counterexample it wakes up the main thread by writing the used seed and size to an \code{MVar}. The main thread will then terminate the remaining testers before it shrinks the counterexample, by delivering asynchronous exceptions. This is very abrupt, with the exceptions delivered at the next allocation point.

\paragraph{graceful}

When a property is aborted as violently as this, by raising an asynchronous exception, there is a risk that there will be artifacts left from a test. If a property e.g. creates a new file on the file system that is normally deleted at the end, an interruption by an asynchronous exception may make the file erroneously persist.

\begin{lstlisting}
prop :: Input -> Property
prop ip = ioProperty $ do
  run $ writeFile "temp.txt" (show ip)
  -- do some work
  run $ deleteFile "temp.txt" -- we may never execute this
\end{lstlisting}

To address this, we introduce a combinator \code{graceful} that takes an \code{IO} action and a handler. The handler will run if QuickCheck makes the choice to terminate evaluation of the property.

\begin{lstlisting}
-- graceful :: IO a -> IO () -> PropertyM IO a

prop :: Input -> Property
prop ip = ioProperty $ do
  run $ writeFile "temp.txt" (show ip)
  graceful
    (do -- do some work
       deleteFile "temp.txt")
    (deleteFile "temp.txt")
\end{lstlisting}

The handler is implemented by intercepting the asynchronous exception before the worker is restarted and running the handler before rethrowing the exception. \textit{graceful} can only capture a specific exception thrown internally by QuickCheck. We choose to implement this dedicated operator like this rather than relying on existing bracket functionality, as both user code and QuickCheck might already have code in place to deal with exceptions.

\code{graceful} can be used not only for shrinking but also for testing. When one tester finds a counterexample the concurrent testers will be aborted. This combinator will make sure that cleanup occurs then as well.

\paragraph{Shrinking}

\begin{figure}
    \begin{subfigure}{0.3\textwidth}
    \centering
        \includegraphics[scale=0.45]{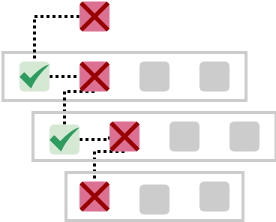}
        \caption{Illustration of the existing QuickCheck shrink-loop. It guarantees to always return the same locally minimal counterexample.}
        \label{fig:existingshrink}
    \end{subfigure}\hfill

    \begin{subfigure}{0.3\textwidth}
    \centering
        \includegraphics[scale=0.45]{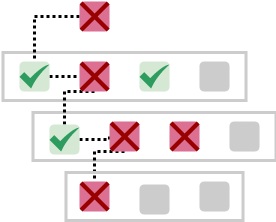}
        \caption{The new deterministic shrink-loop promises to find the same local minimum every time, but it may speculatively evaluate other candidates in its search for the final counterexample.}
        \label{fig:deterministicshrink}
    \end{subfigure}\hfill

    \begin{subfigure}{0.3\textwidth}
    \centering
        \includegraphics[scale=0.45]{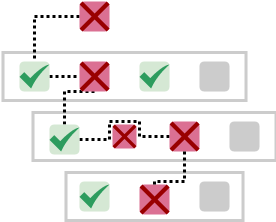}
        \caption{The greedy shrink-loop does not guarantee to find the same local minimum, potentially returning a different final counterexample.}
        \label{fig:greedyshrink}
    \end{subfigure}
    \caption{The three figures above illustrate how the search for a minimized counterexample happened. The dotted line represents the final path to the local minimum, green boxes are candidate counterexamples that turned out to not be new counterexamples, and red boxes are counterexamples that still falsified the property. Grey boxes are candidate counterexamples that were never evaluated.}
    \label{fig:shrinking}
\end{figure}

The existing shrink loop continually evaluates the head of the candidate list until a new counterexample is found, at which point the loop recurses, or until the list is empty, at which point shrinking is terminated. This is illustrated in figure \ref{fig:existingshrink}. The design of the new loop is very similar.

Rather than a single thread traversing the candidate list one element at a time, the parallel shrink loop spawns concurrent worker threads that cooperate and traverse the same list, now residing in an \code{MVar}. If any of the concurrent workers finds a new counterexample, they will update the shared list of candidates and signal to their sibling workers that they should stop evaluating their current candidate and instead pick a new one from the new list.

The behavior of this shrink loop might return a non-deterministic result. Whereas the previous loop will always find the first counterexample in the candidate list, the parallel loop might find a counterexample other than the first one. To emulate the deterministic behavior, the new loop can choose to only signal a restart to those concurrent workers that are evaluating candidates that appeared after the current one in the candidate list, and tell them to speculatively start shrinking the new counterexample. The other workers will keep evaluating their current candidates, and if one of them turns out to be a counterexample, the current progress will be discarded, and shrinking will continue with the new counterexample. In this case, we might do some unnecessary work, but we will get the same deterministic result. Figure \ref{fig:deterministicshrink} illustrates this and how this approach may make us evaluate candidates that we don't need.

Another alternative is that when any worker has found a counterexample, all concurrent workers are restarted and told to start shrinking the new counterexample, regardless if this was the first counterexample or not. This might lead to a non-deterministic result, as the path down the rose tree of shrink candidates is not the leftmost one, as illustrated in figure \ref{fig:greedyshrink}. Restarting a worker is done by raising an asynchronous exception in the worker. The worker will catch this exception and enter the shrink-loop anew, and begin to search through the new list of candidates.

Repeatedly accessing a shared resource may incur overhead costs. If two workers attempt to modify a shared resource at the same time, one will have to wait for the other. As the list of candidate counterexamples is shared between workers, if candidates are evaluated very fast, it is likely that using more threads will slow down shrinking.

\section{Evaluation}

We evaluate QuickerCheck to answer the following four questions

\begin{itemize}
    \item \textbf{Question 1}: Is the sequential performance of the new implementation comparable with QuickCheck?
    \item \textbf{Question 2}: How does the parallel run-time scale as we add more cores?
    \item \textbf{Question 3}: Can we find bugs faster by using more cores?
    \item \textbf{Question 4}: Can we shrink counterexamples faster by using more cores?
    \item \textbf{Question 5}: Does the choice of shrinking algorithm affect the quality of shrunk counterexamples?
\end{itemize}

To answer these questions we run properties and collect information. We will refer to such properties as benchmarks, and the benchmarks we use are described in the following subsection.

\subsection{Benchmarks}

We perform all our evaluations using six distinct benchmarks. While the first benchmark \textit{constant} is artificial, the other benchmarks are intended to represent a diverse set of testing tasks. \textit{compiler testing} and \textit{compressid} are effectful tasks making use of IO facilities, while the other tasks are pure.

\paragraph{constant}

The benchmark named \textit{constant} is not one that anyone would write organically, but its inclusion as a benchmark in this set has a very specific purpose. The underlying property is

\begin{lstlisting}
prop_constant :: () -> Bool
prop_constant () = True
\end{lstlisting}

\begin{figure}
    \centering
    \includegraphics[scale=0.2]{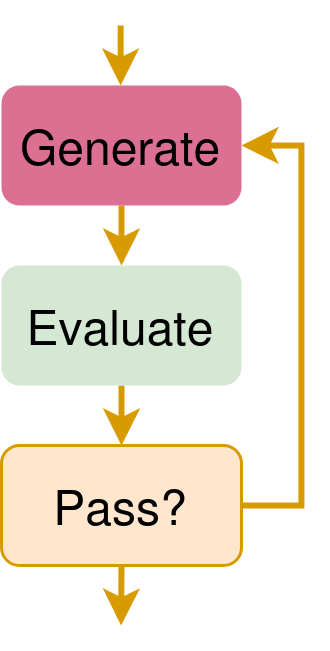}
    \caption{A high-level description of the internal testing loop. The loop begins by generating input and then invoking the property. After this, the loop inspects the outcome before it either reports having found a counterexample, or loops back to repeat all steps. The bottom box and all arrows are part of the internal testing loop, while the top two boxes are defined by the user.}
    \label{fig:testing-loop}
\end{figure}

The cost in execution time of running a test consists of three parts -- generating input, running the property, and the machinery of the internal testing loop. This is illustrated in figure \ref{fig:testing-loop}. As QuickerCheck only changes the workings of the testing loop, we want to measure the change in cost of just the testing loop. The above property minimizes the execution time of both the generation of test data and evaluation of the property. Generation and evaluation are constant time as there are no random choices to make during generation and evaluation of the property is trivial. Measured changes in the execution speed of QuickCheck vs QuickerCheck on this benchmark should primarily be a result of the different testing loops.

\paragraph{compiler testing}

The underlying property of the \textit{compiler testing} benchmark asserts that a compiler for an imperative language generates correct output. The property is stated as a metamorphic relation as described in section \ref{para:compilertesting}.

In practice, the property does significantly more work than the other benchmarks. It generates a type-correct imperative program and produces several executables that are invoked to assert the correctness. The generated programs may include non-terminating loops, so the property might require some time to execute. Such loops are eventually broken by the property itself after consuming too many resources. During evaluation, the property will spend a significant amount of time in external processes.

\paragraph{compressid}

This benchmark composes the two Unix commands \textit{gzip} and \textit{gunzip} and verifies that the composition behaves as the identity function. It generates an arbitrary string and invokes \code{gzip}, passes the compressed result to \code{gunzip}, and asserts that the final output is identical to the input.

This benchmark comes in three flavors -- one is a naive implementation (\textit{naive}) that writes intermediary values directly to the file system. Since the file system is a shared resource a property like this will experience race conditions if multiple threads are used. We test two alternative implementations that make the property thread-safe in different ways. The first (\textit{tmpfs}) generates fresh directories for each concurrent worker to write such files to, and the second (\textit{nofs}) uses pipes to pass values around, never using the file system.

\paragraph{verse}

This property asserts the confluence of the rewrite system for the Verse Core Calculus \cite{verse}. A rewrite system is confluent if, regardless of which rewrite rules are applied in each step, the result is always the same, single, normal form.

The property generates an arbitrary term and applies two arbitrary sequences of rewrite rules. If the two resulting normal forms are different, the rewrite system is not confluent and the property is falsified.

\paragraph{system f}

The \textit{system f} benchmark is a pure property that generates arbitrary lambda terms and asserts the subject reduction property, described in section \ref{sec:introduction}, which states that the type of a term should not change after performing one reduction of said term. The code was taken from Etna, an evaluation platform for Property-based testing frameworks\cite{shi2023etna}.

\paragraph{twee}

Twee \cite{DBLP:conf/cade/Smallbone21} is a high-performance theorem prover for equational logic written in Haskell. A key component is the \emph{term index}, a data structure for finding equations matching a given term. The \textit{twee} benchmark is a pure property stating that, after any sequence of update operations on a term index, the data structure's invariant is preserved.

\subsection{Results and Discussion}

Evaluation is done using an Intel I7-10700 8-core CPU with turbo-boost turned off. The evaluation system is equipped with 64GB of 2933MT/s RAM.

\sloppy
We use GHC to compile and execute Haskell code, using the compile-time flags \code{-threaded}, \code{-feager-blackholing}, and \code{-rtsopts}. We don't try to mitigate garbage collection costs by increasing the nursery size or try to improve the performance in any other way, as we believe most people use QuickCheck without doing this. All invocations of QuickCheck are made with the \code{chatty} flag set to \code{False} as printing would otherwise affect the results. In appendix \ref{ap:chatty} it is illustrated how \code{chatty} affects experimentation.

\fussy
\paragraph{Is the sequential performance of the new implementation comparable with QuickCheck?}

\begin{figure}
    \centering
    \includegraphics[scale=0.5]{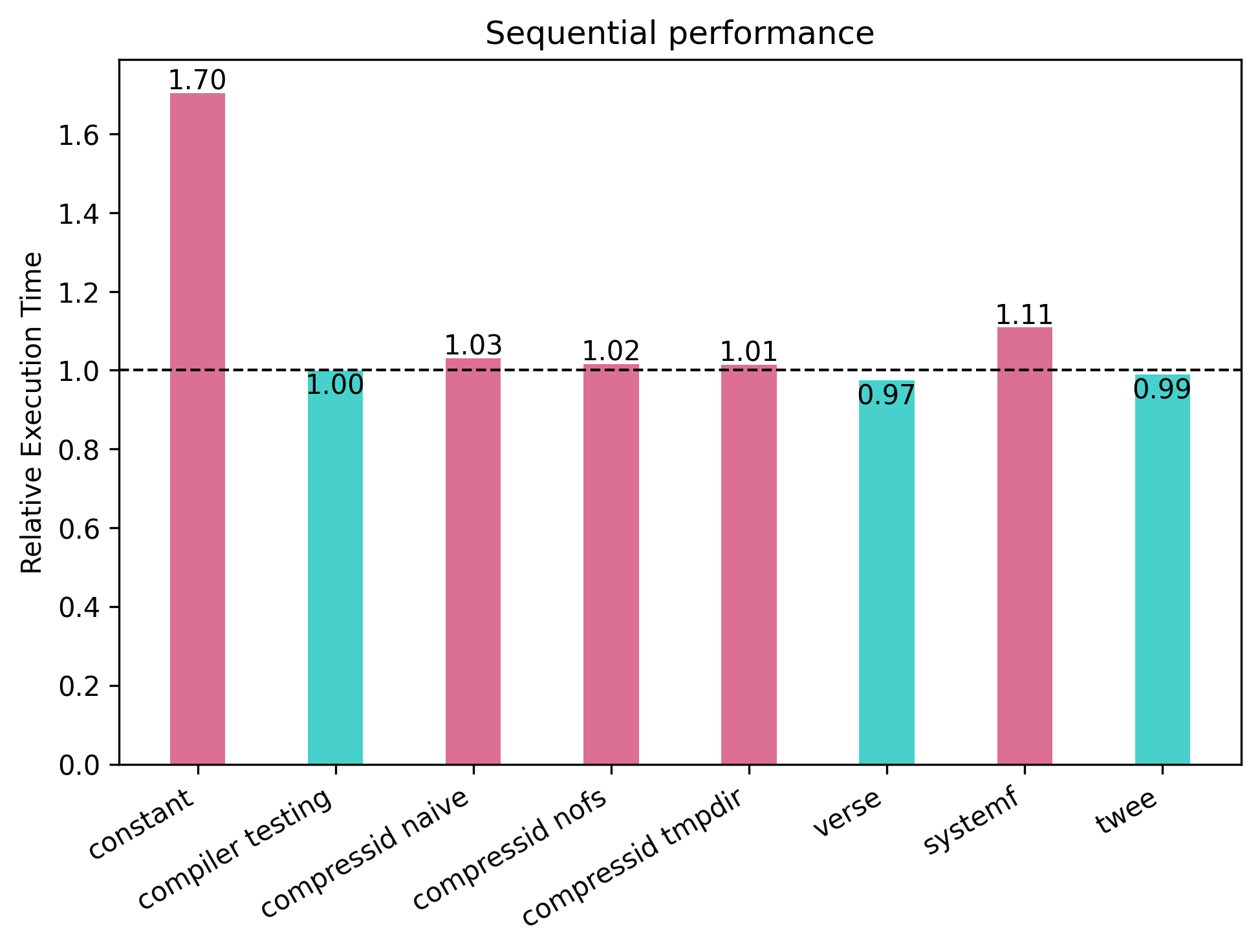}
    \caption{The performance of sequential QuickerCheck compared to that of QuickCheck. A number of 1 means that there was no difference in performance, whereas a number less than 1 indicates that QuickerCheck was faster than QuickCheck (e.g. 0.5 shows that QuickerCheck finished in half the time). A number greater than 1 indicates that QuickerCheck was slower than QuickCheck.}
    \label{fig:seqperf}
\end{figure}

We answer this by executing each benchmark several times both with QuickCheck and QuickerCheck, using only one core. We compute the median execution times and compare them. The results are presented in figure \ref{fig:seqperf}.

Something that immediately stands out is the huge overhead experienced by the \textit{constant} benchmark. This benchmark is intended to act as a worst-case property and illustrate precisely what the overhead of the new testing loop is. The results indicate that, in the worst case, QuickerCheck will incur a penalty of 70\%.

The other benchmarks all perform some actual workload and experience much more modest changes in performance. The \textit{system f} property, just like the \textit{constant} property, is very fast. By running many more tests it interacts much more with the new testing loop, incurring more of the new costs. This shows up by QuickerCheck requiring 11\% more execution time to finish the same workload. Some of the workloads experienced no change at all or even got slightly faster.

Not accounting for the \textit{constant} benchmark, it appears that there is no major change in performance by using sequential QuickerCheck instead of QuickCheck.

\paragraph{How does the parallel run-time scale as we add more cores?}

\begin{figure}
    \centering
    \includegraphics[scale=0.59]{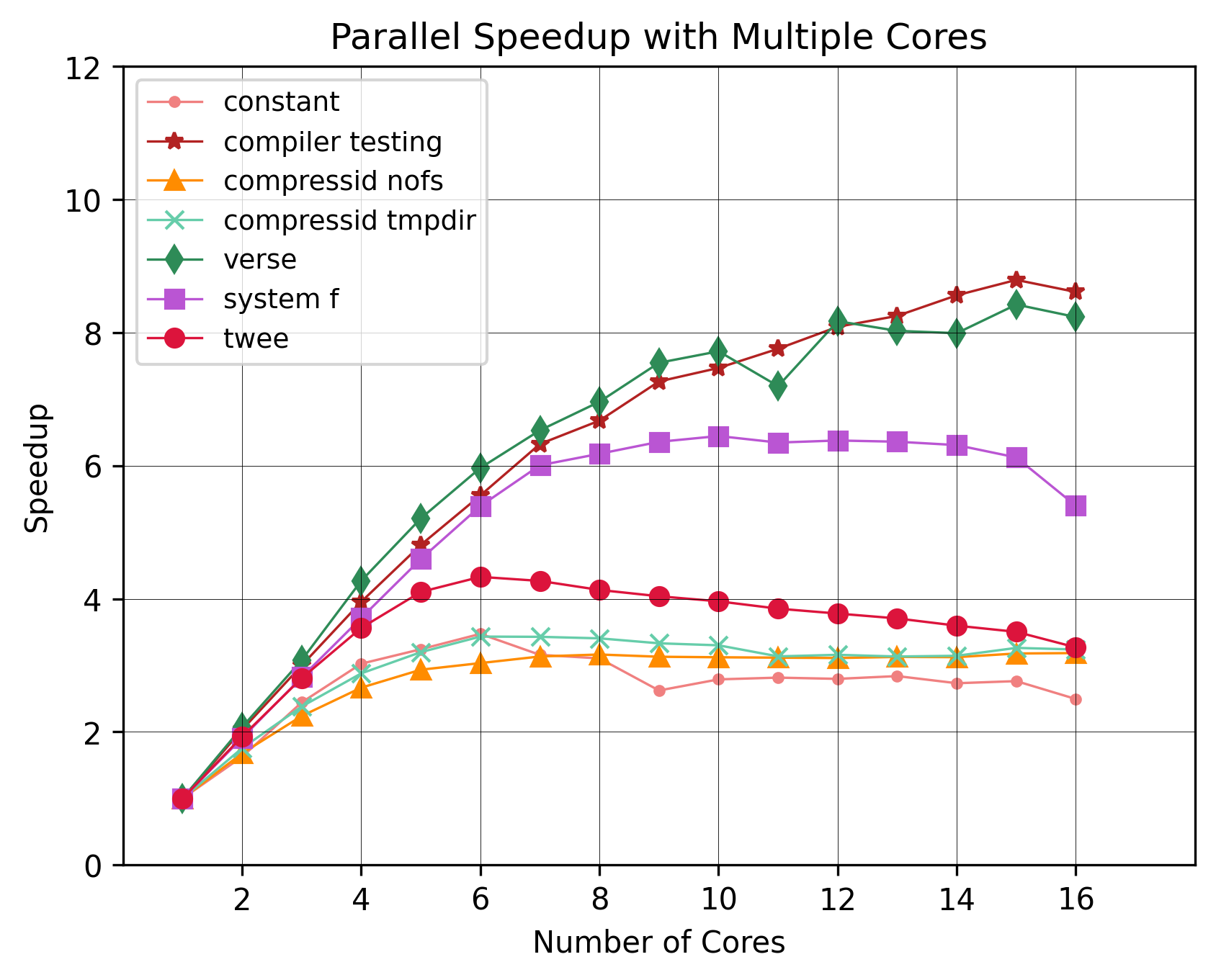}
    \caption{The acquired speedup relative to the sequential execution time when running tests.}
    \label{fig:speedup}
\end{figure}

Each of the benchmarks is executed several times for each core configuration, the median execution time is computed and the speedup relative to the sequential running time is computed. The results are presented in figure \ref{fig:speedup}.

We first observe that many of the benchmarks scale very well until the point where we exhaust the number of physical cores. The highest speedup was achieved by the \textit{compiler testing} benchmark which got more than eight times faster. The \textit{verse} property is not far behind.

The \textit{twee} benchmark initially scales very well but starts to lose momentum when we approach the limit of physical resources. When hyper-threading is active performance slowly but surely degrades. The \textit{twee} benchmark is very data-intensive and frequently moves data around. While two hyper-threads appear to the operating system as two CPUs, they are actually two logical threads that share hardware components required to execute machine instructions, such as caches and the system bus. One potential explanation for this degradation is that the different testers affect the cache in unfavorable ways.

One noticeable difference between e.g. the \textit{compiler testing} and \textit{system f} benchmark is that the \textit{compiler testing} property is significantly slower. The property may spend over a second evaluating a single test while the \textit{system f} benchmark may run thousands of tests in the same time frame. The results seem to indicate that the more time a property spends inside the body of the property, the greater the potential speedup.

\paragraph{Can we find bugs faster by using more cores?}

\begin{figure}
    \centering
    \includegraphics[scale=0.59]{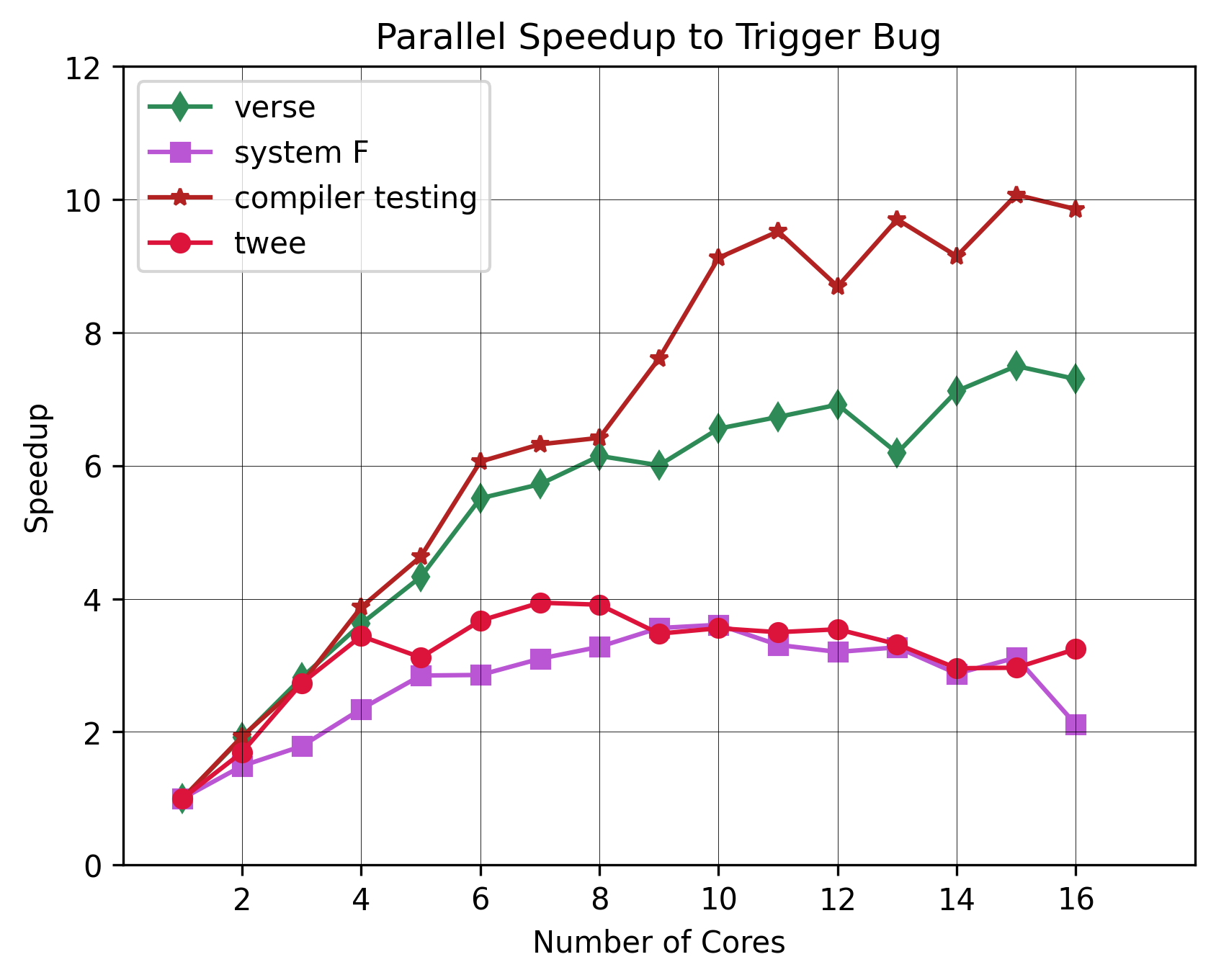}
    \caption{The acquired speedup relative to the sequential execution time when searching for a planted bug.}
    \label{fig:bugs}
\end{figure}

To evaluate this we plant a bug in 4 of the 6 benchmarks and let QuickerCheck run until it finds the bug. This is repeated 300 times after which the median execution time is computed. Figure \ref{fig:bugs} illustrates the speedup acquired relative to the sequential execution time.

We first note that the \textit{system f} benchmark doesn't reach as high of a speedup as when we are running tests without a bug enabled. While we can't say with certainty what to attribute this to, we have a pretty good guess of what is happening. The shape of the curve is the same, except that it is pushed down towards lower multiples. There are some new costs associated with starting up the parallel test loop, and when we run tests without a bug enabled the benchmark is allowed to run for a few seconds, running hundreds of thousands of tests. The cost of starting up the test loop is amortized over all these tests, while when a bug is enabled there are many fewer tests. The bug was found after roughly 200 tests, running for just a couple of milliseconds.

The overall shape of the \textit{twee} benchmark is the same, but not reaching quite as high of a speedup as when just running tests. The \textit{compiler testing} benchmark acquires a 10x increase in performance, outperforming all other evaluated benchmarks. We believe this speedup is higher than that achieved in figure \ref{fig:speedup} because many concurrently running tests are aborted when a counterexample is found. When evaluating speedup for question two, every test that we began evaluating was expected to finish, whereas when we evaluated question three, we terminated concurrent testers when one of them found a counterexample. We will thus do slightly less work. We observe the inverse behavior in the \textit{system f} property, where the concurrent testers have time to run many additional tests before they are terminated by a tester who found a counterexample.

\paragraph{Can we shrink counterexamples faster by using more cores?}

We generate 200 random seeds that we know trigger bugs, such that we can replay them to deterministically see the same counterexamples. We replay the seeds and measure how long it takes to shrink them, varying the number of cores and the choice of strategy (deterministic or greedy shrinking). We have done this for the three benchmarks \textit{compiler testing}, \textit{twee}, and \textit{verse}. Because it is impractical to show all the results, we have picked some subsets of data that we find representative of the overall results.

\begin{figure}
    \centering
    \includegraphics[scale=0.5]{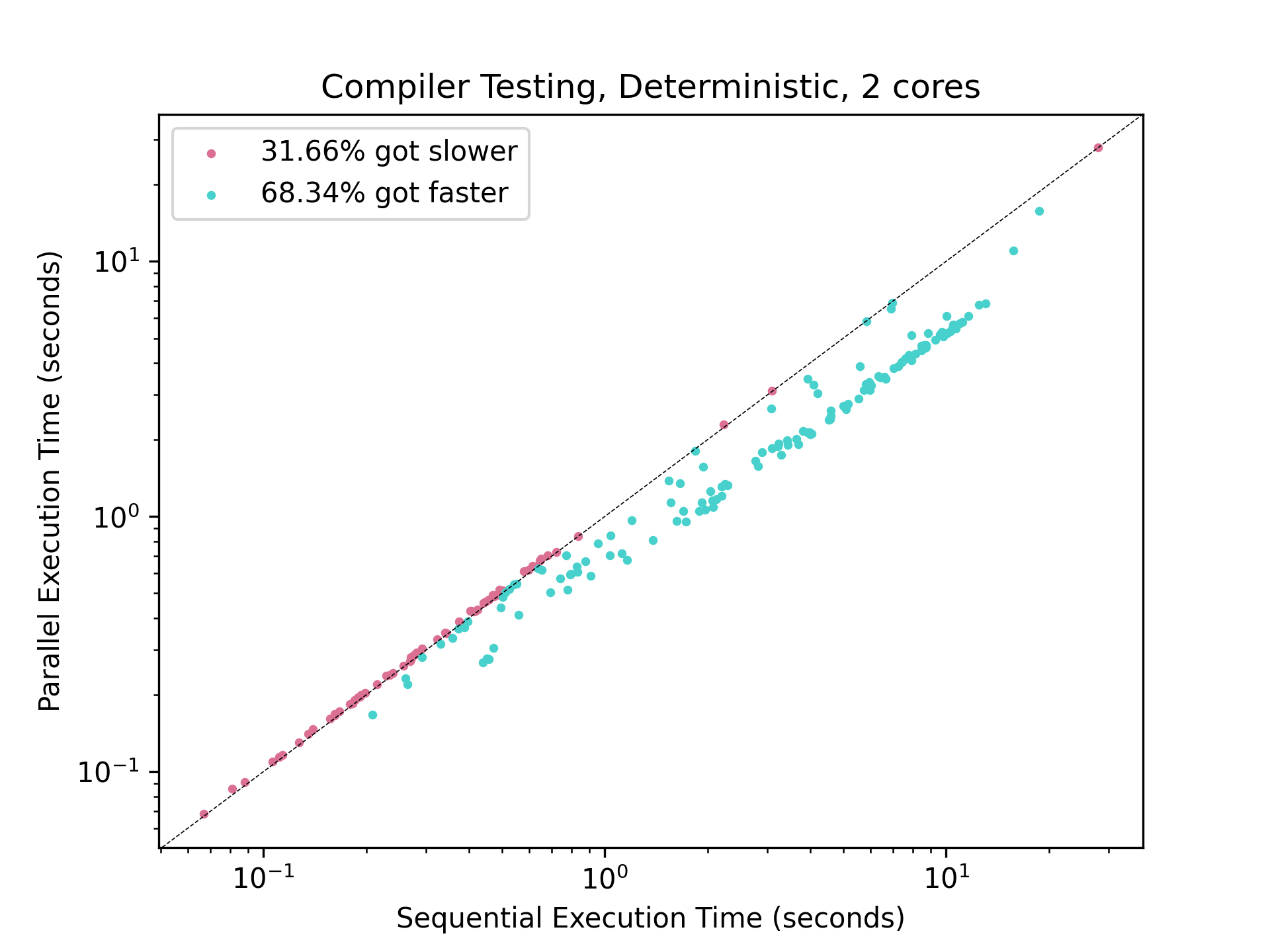}
    \caption{The speedups acquired when using two cores to shrink the \textit{compiler testing} tests, using the deterministic algorithm.}
    \label{fig:deterministic-comptimes-2-ic}
\end{figure}

\begin{figure}
    \centering
    \includegraphics[scale=0.5]{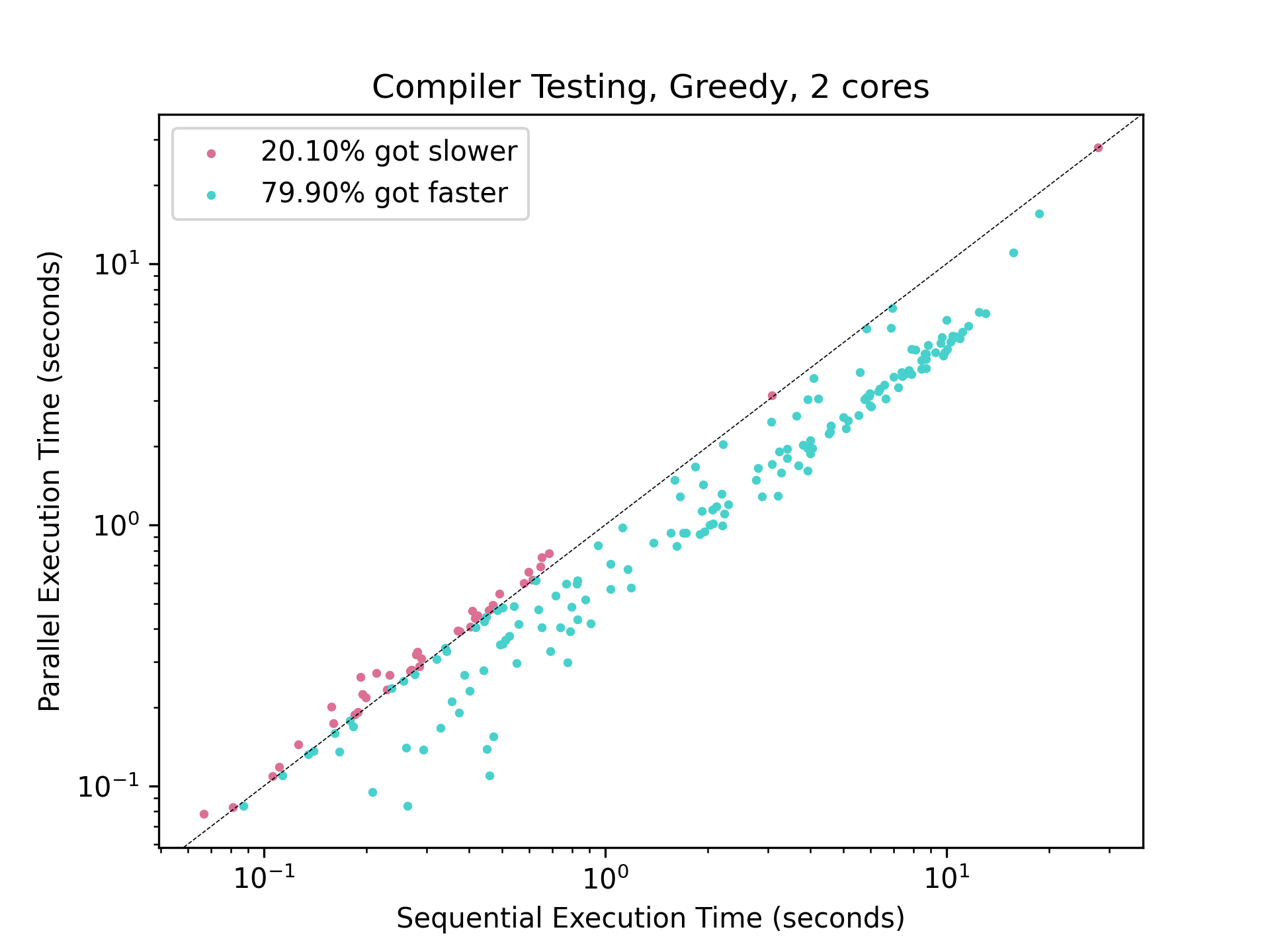}
    \caption{The speedups acquired when using two cores to shrink the \textit{compiler testing} tests, using the greedy algorithm.}
    \label{fig:greedy-comptimes-2-ict}
\end{figure}

\begin{figure}
    \centering
    \includegraphics[scale=0.5]{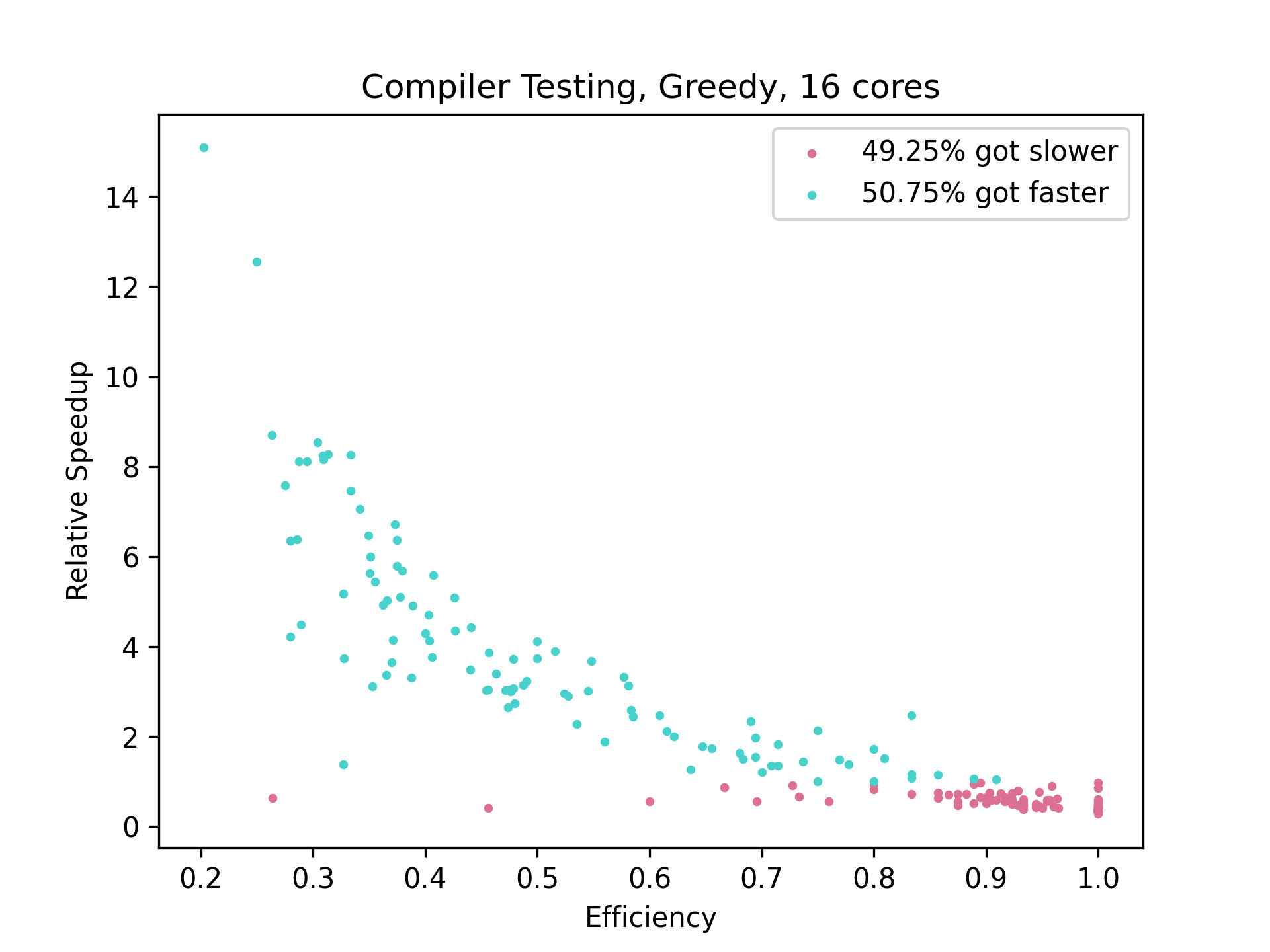}
    \caption{This figure illustrates that the closer the efficiency is to one, the higher the probability that the test will get slower when shrinking. It also appears that the relative speedup is higher the lower the efficiency.}
    \label{fig:greedy-shrink-efficiency-ict}
\end{figure}

The \textit{compiler testing} results are presented in figure \ref{fig:deterministic-comptimes-2-ic}, \ref{fig:greedy-comptimes-2-ict}, and \ref{fig:greedy-shrink-efficiency-ict}. Figures \ref{fig:deterministic-comptimes-2-ic} and \ref{fig:greedy-comptimes-2-ict} illustrate the relationship between sequential and parallel execution time, using two cores. The red dots got slower when two cores were used, whereas the blue dots achieved a speedup. The further from the line a point lies, the more extreme the achieved effect is. From the two figures, we can see that the greedy algorithm appears to benefit more experiments and that the achieved effects are greater. The blue dots in figure \ref{fig:deterministic-comptimes-2-ic} appear to tangent a line. This line traces the execution time that is twice as fast as the sequential one and illustrates the upper bound defined by Amdahl's law\cite{DBLP:conf/afips/Amdahl67}. The results in figure \ref{fig:greedy-comptimes-2-ict} show some experiments crossing this boundary, which is explained by the greedy algorithm being able to return a completely different counterexample.

As more and more cores are added, the results indicate that more and more experiments got slower, while the remaining ones that achieved a speedup achieved a much greater speedup.

To try and answer which counterexamples may benefit from parallel shrinking, we plot the \textit{efficiency} of the shrunk counterexamples. The efficiency of a single counterexample is defined as the fraction of evaluated candidates that successfully shrunk the counterexample, and as such is a number between 0 and 1. It is clear that if the efficiency is one, there is nothing to be gained from parallelism as shrinking becomes a sequential search. As an example, the total number of evaluated candidates in figures \ref{fig:existingshrink}, \ref{fig:deterministicshrink}, and \ref{fig:greedyshrink} are 5, 7, and 8 respectively. In all 3 cases the number of successful shrinks was 3, so the efficiencies are 0.6, 0.42, and 0.375.

Figure \ref{fig:greedy-shrink-efficiency-ict} shows that there is a clear trend of counterexamples with a good efficiency not benefiting from parallel shrinking. If there was not that much extra work to be done from the beginning, the existence of more cores does not offer any substantial performance improvements.

\begin{figure}
    \centering
    \includegraphics[scale=0.5]{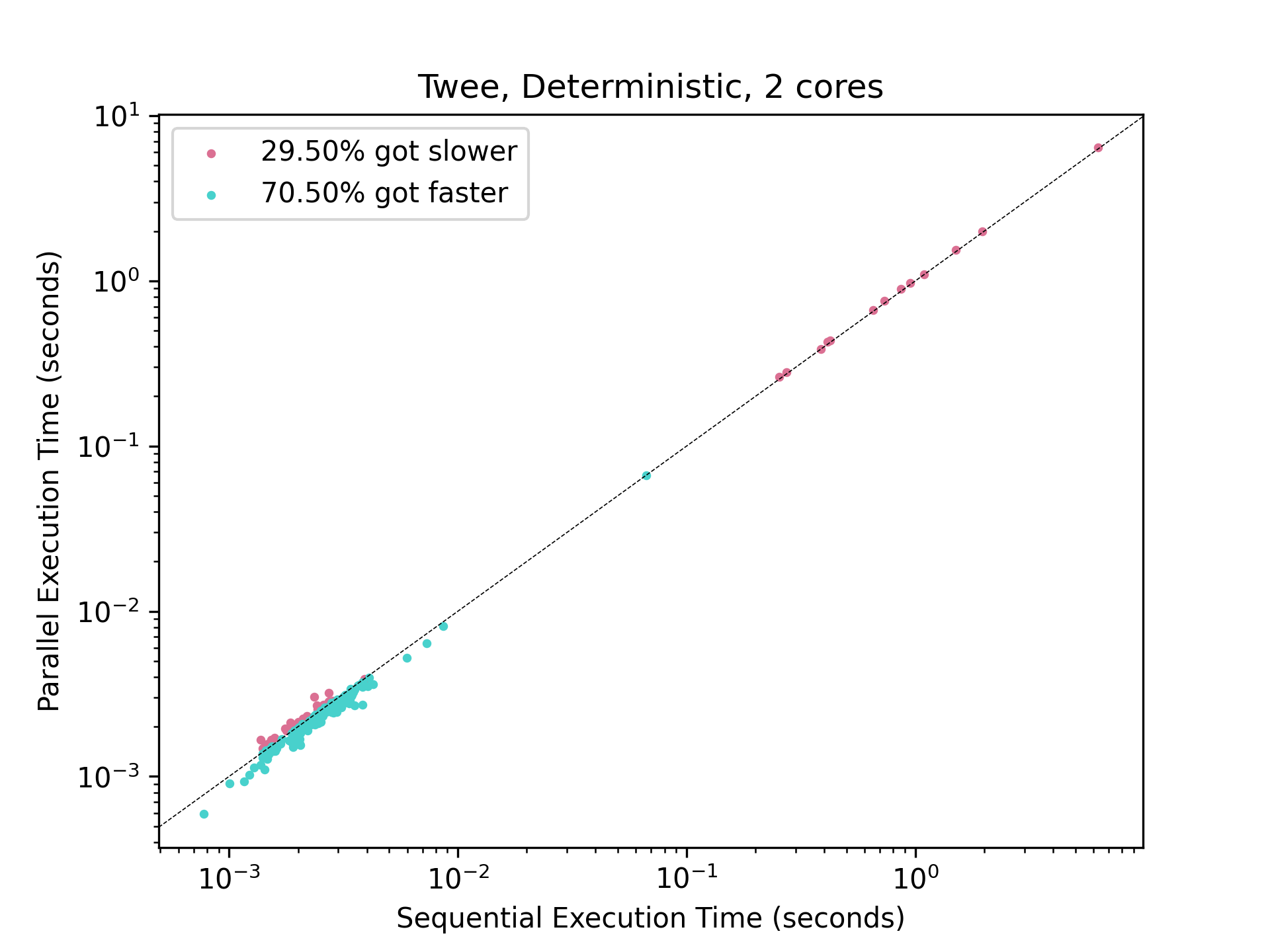}
    \caption{The results indicate that most tests finished shrinking very fast when only two cores was used.}
    \label{fig:deterministic-comptimes-2-twee}
\end{figure}

\begin{figure}
    \centering
    \includegraphics[scale=0.5]{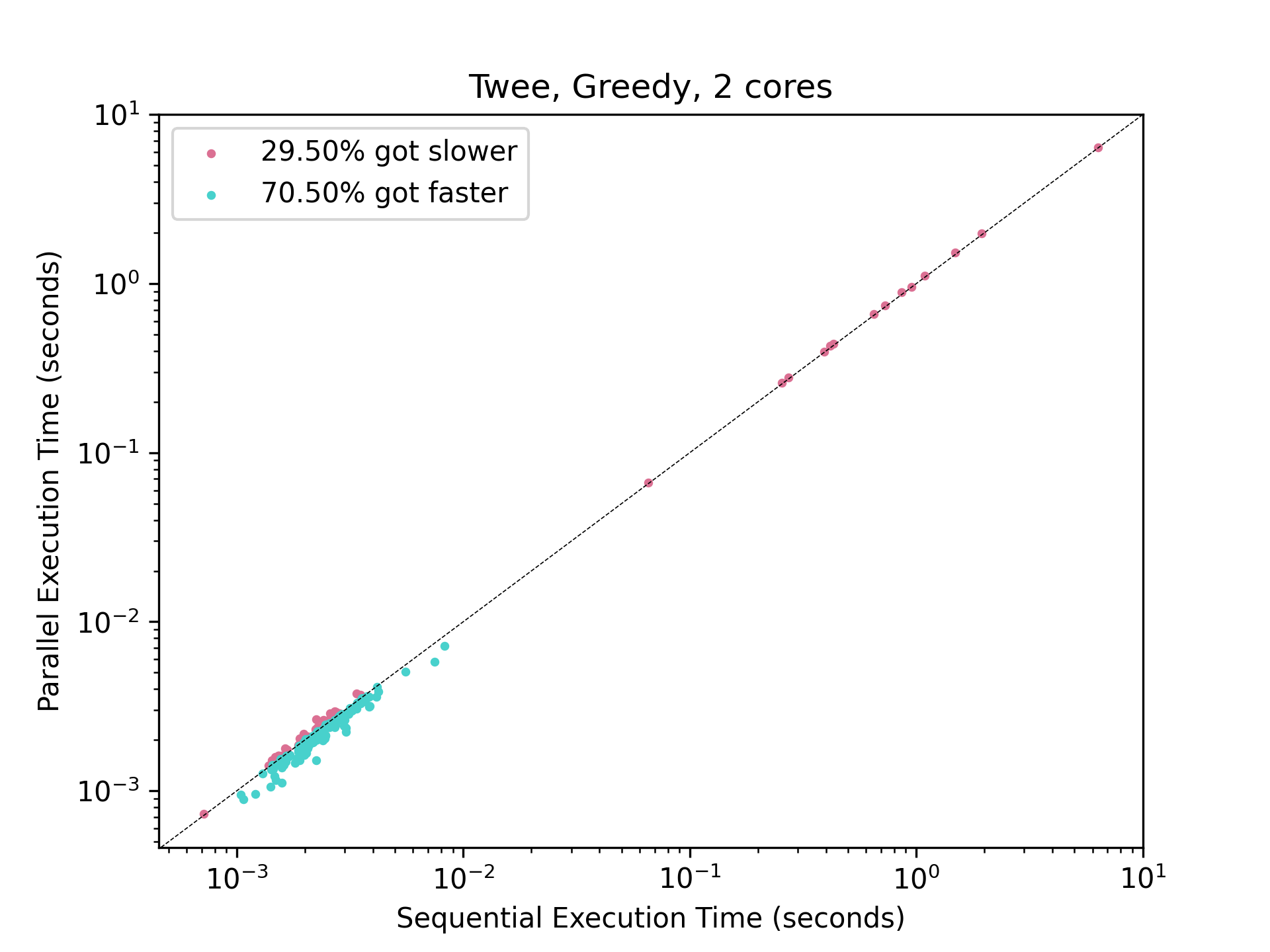}
    \caption{The greedy algorithm appears to perform roughly the same as the deterministic one, with the exception of some tests that did indeed shrink faster.}
    \label{fig:greedy-comptimes-2-twee}
\end{figure}

\begin{figure}
    \centering
    \includegraphics[scale=0.5]{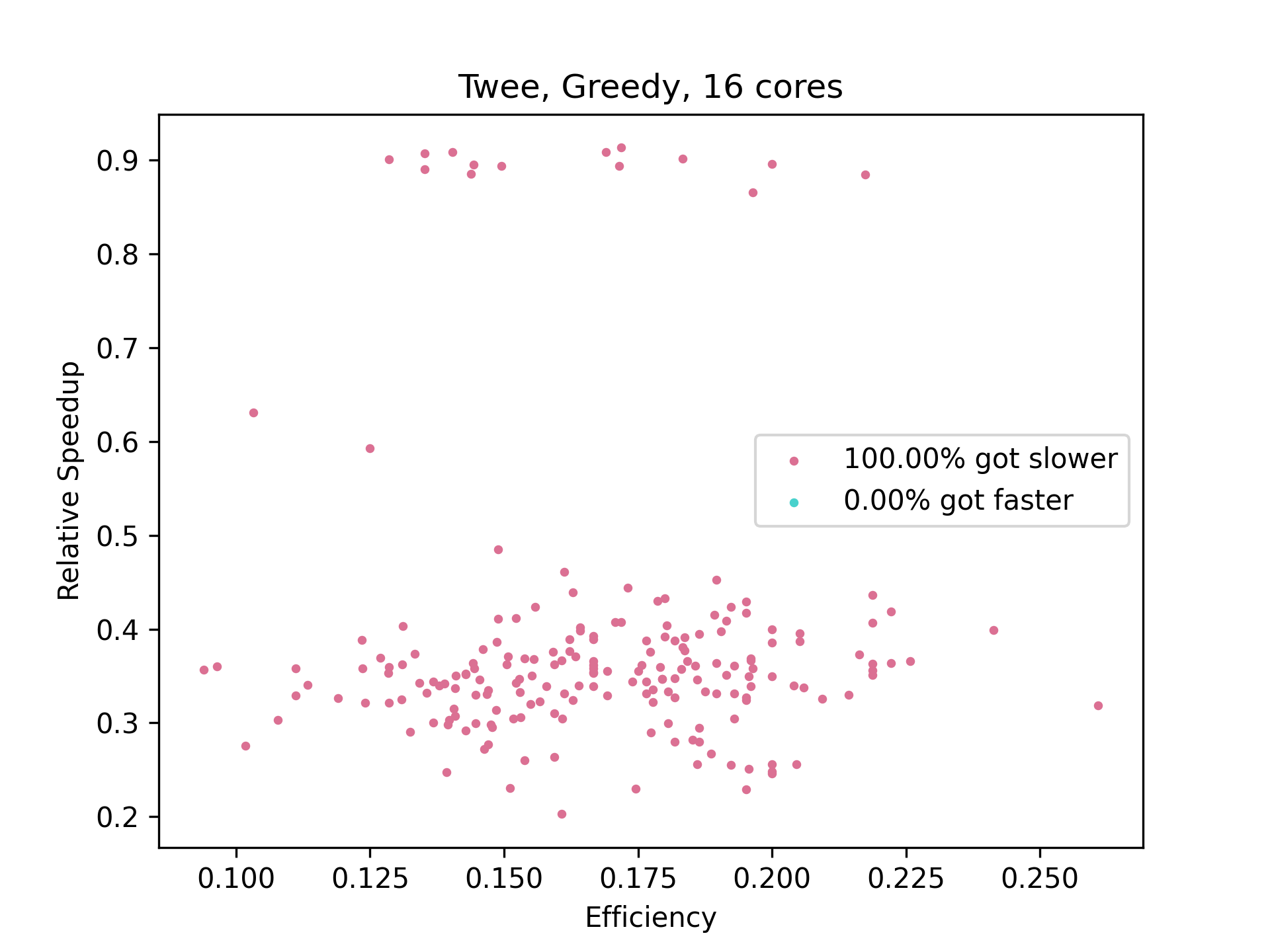}
    \caption{While the efficiency turned out to be an excellent indicator for whether a test got faster or not for the \textit{compiler testing} benchmark, the same can not be said for \textit{twee}. Using 16 cores and the greedy algorithm, all tests got slower and there was quite a spread of efficiencies. The overall efficiency appears to be much lower, but there is still nothing to be gained by additional cores.}
    \label{fig:greedy-shrink-efficiency-twee}
\end{figure}

The results observed from \textit{twee} (figures \ref{fig:deterministic-comptimes-2-twee}, \ref{fig:greedy-comptimes-2-twee}, and \ref{fig:greedy-shrink-efficiency-twee}) tell a different story. The \textit{twee} property finishes shrinking in a couple of milliseconds, and using more cores quickly makes all observed counterexamples shrink slower. The efficiency appears to make no difference and we believe that the overhead of the parallel search overshadows any benefits of using more cores. The advantage of having more cores at one's disposal appears to mainly be beneficial in cases where execution will require a non-trivial amount of time.

\begin{figure}
    \centering
    \includegraphics[scale=0.5]{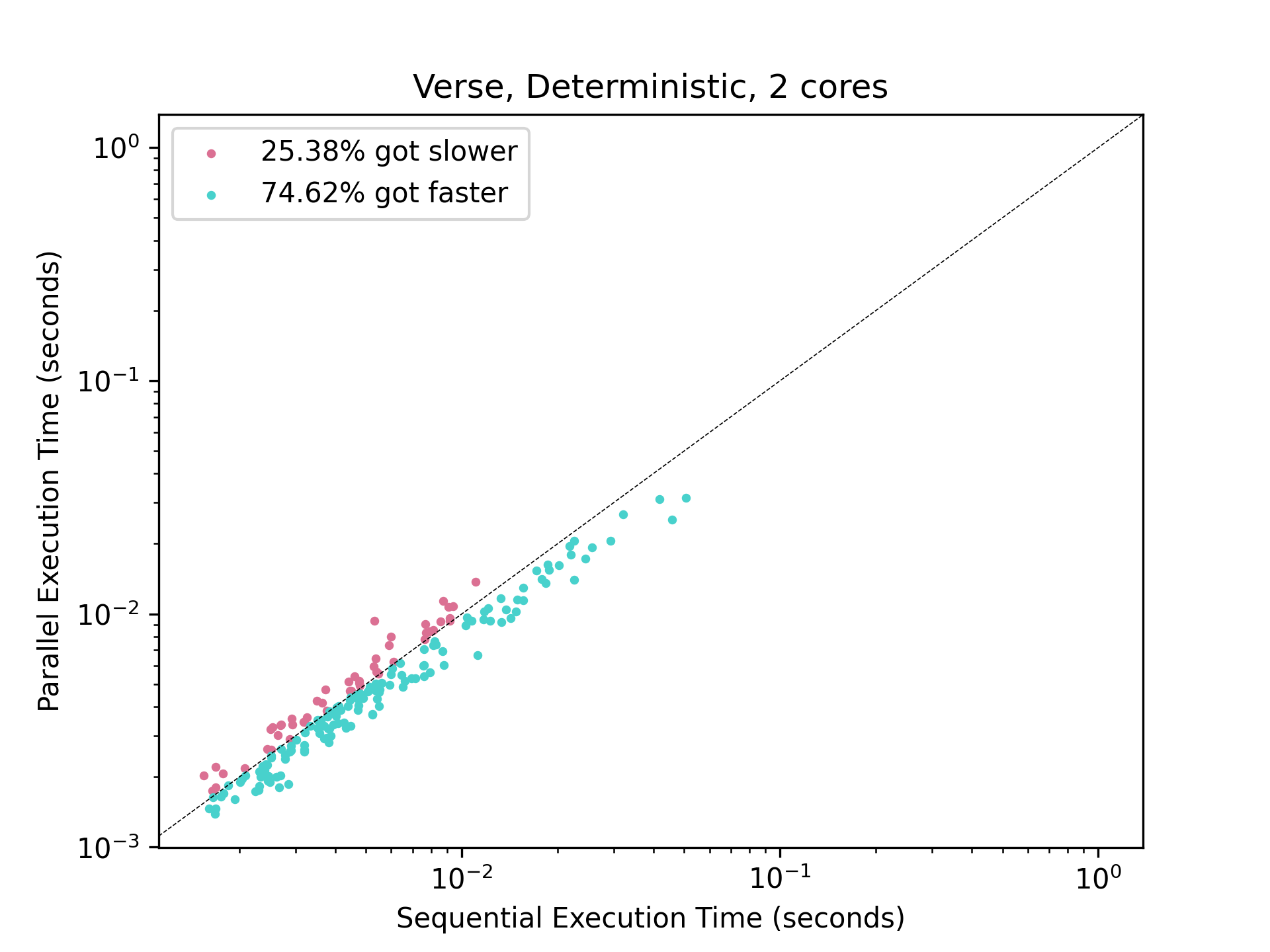}
    \caption{The speedups acquired with two cores using the deterministic algorithm, for the \textit{verse} benchmark.}
    \label{fig:det-shrink-verse}
\end{figure}

\begin{figure}
    \centering
    \includegraphics[scale=0.5]{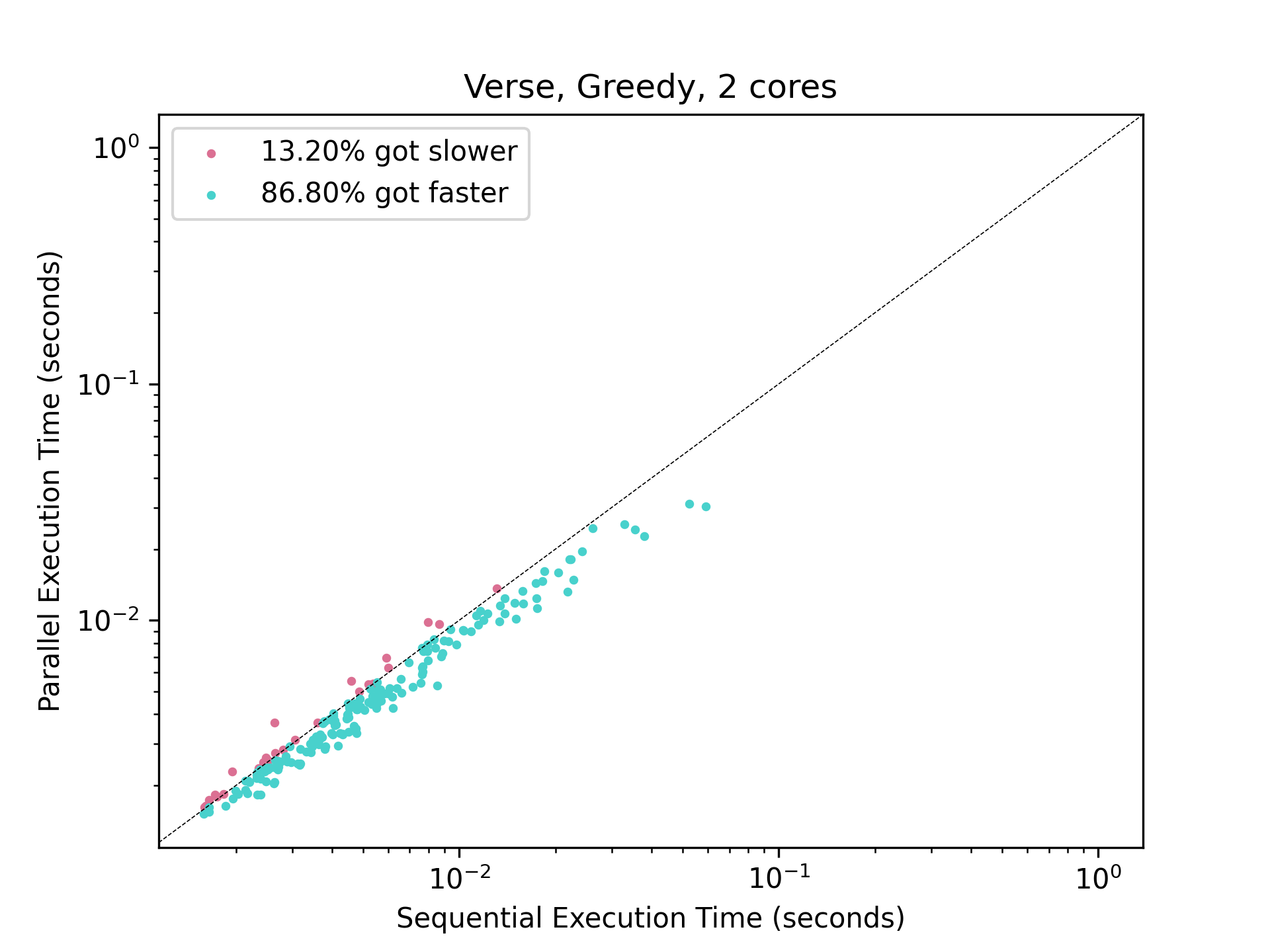}
    \caption{The speedups acquired with two cores using the greedy algorithm, for the \textit{verse} benchmark. It can be observed that the number of candidates that achieved a speedup increased, compared to using the deterministic algorithm.}
    \label{fig:gree-shrink-verse}
\end{figure}

\begin{figure}
    \centering
    \includegraphics[scale=0.5]{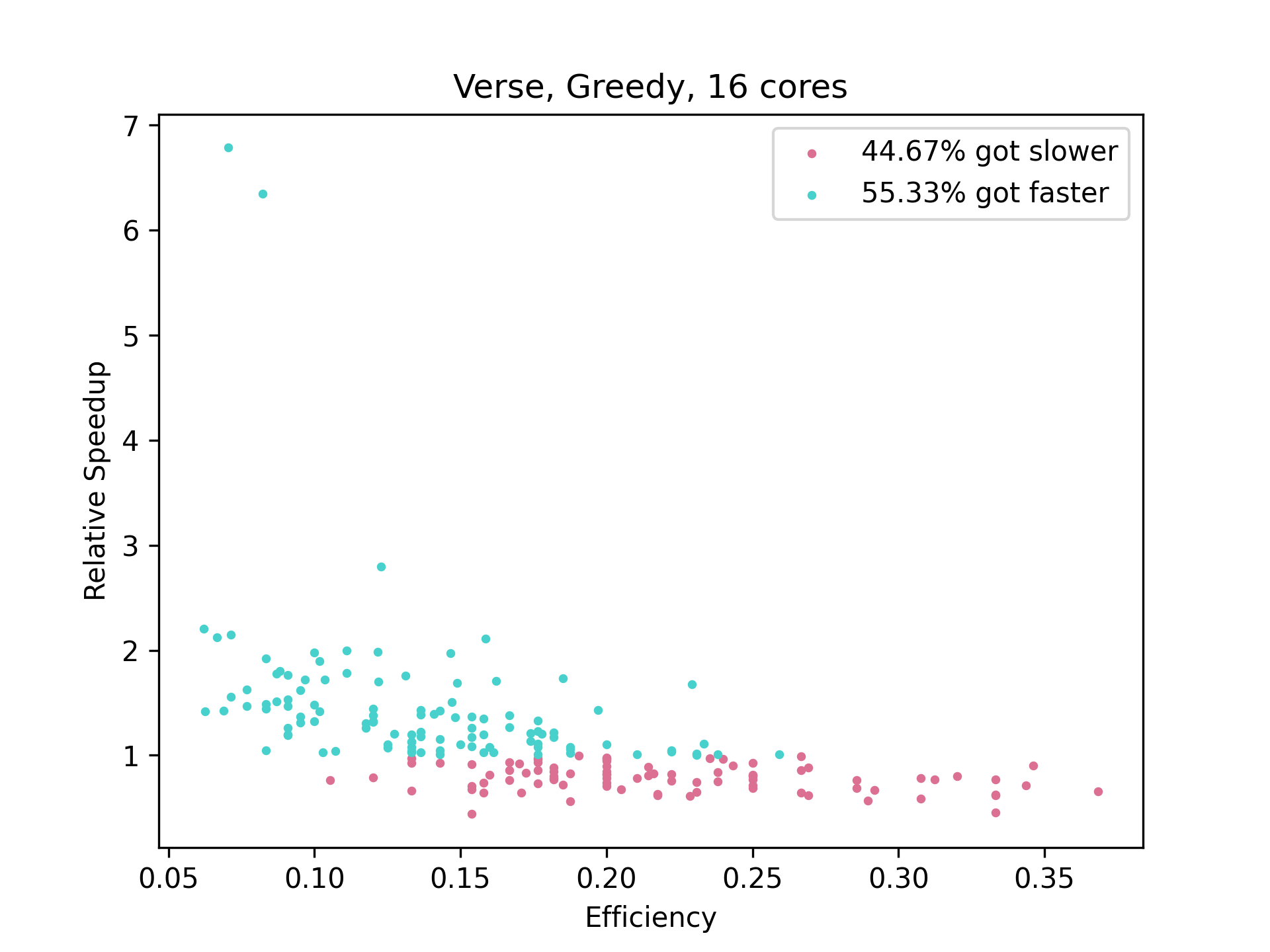}
    \caption{The efficiency of the \textit{verse} shrinker shows that there is a slight trend of lower efficiency indicating that there is a speedup to have by using more cores.}
    \label{fig:gree-shrink-efficiency-verse}
\end{figure}

The \textit{verse} benchmark, much like the \textit{compiler testing} one, achieves a noticeable speedup for the majority of candidates. This is illustrated in figures \ref{fig:det-shrink-verse} and \ref{fig:gree-shrink-verse}. The efficiency of the shrinker is depicted in figure \ref{fig:gree-shrink-efficiency-verse}, and shows that there is a slight trend towards candidates with a lower efficiency being more likely to benefit from multiple cores. This benchmark shrinks quite rapidly, and as we add more cores, more and more candidates become slower, with the final number at 16 cores showing that roughly half of the candidates experienced a slowdown.

\paragraph{Does the choice of shrinking algorithm affect the quality of shrunk counterexamples?}

The deterministic shrinking algorithm will always yield the same locally minimal counterexample, while the greedy algorithm may return another local minimum, of a potentially different size. We are interested in finding out whether the distribution of sizes of shrunk counterexamples is different for the two algorithms. We evaluate this on two benchmarks, \textit{compiler testing} and \textit{verse}. We collect 300 seeds from the \textit{compiler testing} benchmark and 500 from the \textit{verse} benchmark. These seeds immediately falsify the property, allowing us to shrink them and record the size using both algorithms. We define the size of a counterexample as the number of constructors in it. We point out that both algorithms produce identical results when only one core is used.

\begin{figure}
    \centering
    \includegraphics[scale=0.5]{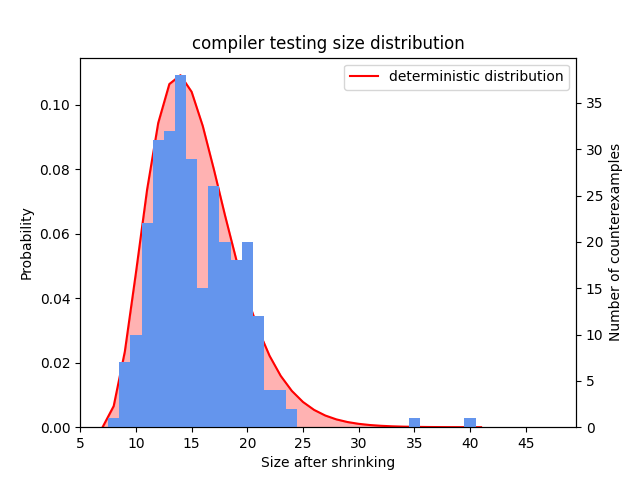}
    \caption{The measured sizes are rendered as a histogram, together with a model that represents the distribution from which they were drawn.}
    \label{fig:with-hist}
\end{figure}

To compare the results from the two algorithms, we model the measured sizes as negative binomial distributions. Whereas we only have one baseline model (the deterministic algorithm), we have 16 models representing the greedy algorithm (one for each core configuration). The authors note that in the single-core case, the two algorithms are identical. Figure \ref{fig:with-hist} illustrates both the measured sizes of the deterministic algorithm, as well as the model representing them.

\begin{figure}
    \centering
    \includegraphics[scale=0.5]{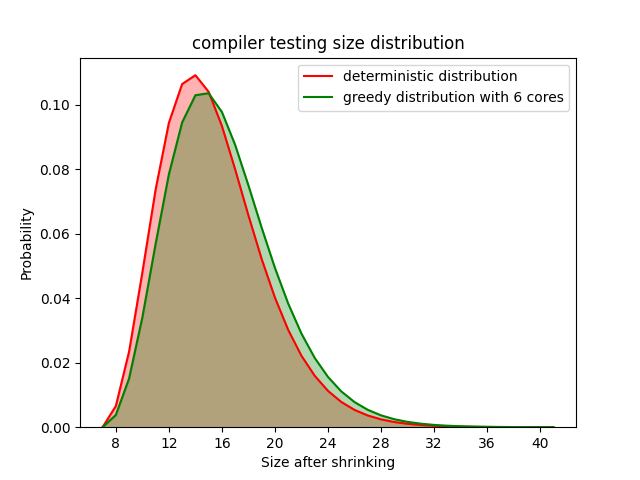}
    \caption{The figure illustrates the distribution of the baseline samples, as well as the greedy distribution with the highest relative entropy, from the \textit{compiler testing} benchmark.}
    \label{fig:compiler-testing-maxdiff}
\end{figure}

\begin{figure}
    \centering
    \includegraphics[scale=0.5]{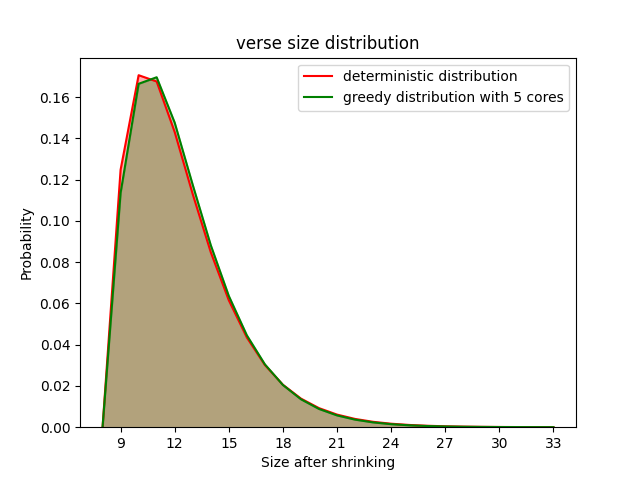}
    \caption{The figure illustrates the distribution of the baseline samples, as well as the greedy distribution with the highest relative entropy, from the \textit{verse} benchmark. The distributions are practically the same.}
    \label{fig:verse-maxdiff}
\end{figure}

We compare the models representing the greedy algorithm to the baseline model by computing the entropy between them. Figures \ref{fig:compiler-testing-maxdiff} and \ref{fig:verse-maxdiff} illustrate the baseline model and the greedy model with the highest relative entropy. In both measured benchmarks the difference is very small. While the \textit{verse} benchmark shows little to no difference at all, the \textit{compiler testing} benchmark has a small but noticeable difference. This difference is not large enough to say whether the distributions are different or not. The results indicate that the choice of algorithm does not impact the quality of shrunk counterexamples at all.

\section{Related Work}

QuickCheck, having proven itself an extremely useful framework for testing software, has been re-implemented in many programming languages. It appears that most other implementations don't support parallel execution of properties. The only other implementation we could find that supports parallelism is \textit{fsCheck}, a QuickCheck implementation for testing \textit{.NET} code. The parallel run-time is not described in any paper and the documentation is sparse, but the implementation is discussed in a merge request introducing the work. The discussion indicates that they initially used an offset to compute sizes for tests but switched to using a stride after observing an uneven workload between workers.

The largest framework for property-based testing by number of users is the Python package \textit{Hypothesis}. They have explicitly chosen not to provide support for parallel evaluation of properties as it can not be determined beforehand whether the function being tested is thread-safe or not. In a non-pure language like Python, this might be a concern, but we believe that this concern is not as severe when it comes to Haskell code. Haskell code is usually split up into its pure parts and effectful parts, with the pure parts being embarrassingly parallel from the get-go. Effectful code can in many cases be refactored to be thread-safe, such that parallel testing may yield positive results.

The Haskell package \textit{tasty} \cite{tasty} lets the user define test suites with individual tests in a suite being of different kinds. A test suite can simultaneously include e.g. QuickCheck tests, SmallCheck tests, and unit tests. This is possible by \textit{tasty} using different test drivers to execute the tests. \textit{tasty} can execute individual tests in a test suite in parallel, but it will not introduce parallelism in the underlying test drivers. If a test suite contains many tests, with all but one test terminating very quickly, the majority of execution time will be sequential, waiting for the longest running test to terminate.

\section{Conclusions and Future Work}

Our results show that parallel testing is beneficial. If the property being tested is slow to run the expected performance increase is high, whereas a fast property stands to gain less (but not nothing).

Thanks to the natural division of effectful and pure code in Haskell, many properties are immediately able to benefit from the parallel run-time. We found that with slight modifications to effectful properties, we could run them in a thread-safe manner.

Parallel shrinking is not as universally beneficial, but can still yield good results. For all benchmarks evaluated, individual counterexamples could go either way, either experiencing a slowdown or a speedup. We can not conclude that parallel shrinking is always beneficial. It depends on not only the property but also the specific test case. As more cores are added, some counterexamples will get significantly faster, while the likelihood of your counterexample shrinking slower increases. There seems to be a good compromise around using multiple cores, but a lower number. The greedy algorithm appears to offer a greater speedup than the deterministic one, without compromising on the quality of the counterexamples.

While the work presented in this paper represents a considerable engineering effort, there are still many lines of future work to pursue. While implementing the work described in this paper, it became clear that the ad-hoc way of computing sizes in QuickCheck does not lend itself nicely to parallelism. It imposes a sequential ordering to test cases and is tricky to distribute over multiple cores. While we have implemented a best-effort attempt to maintain the previous behavior, it is not a perfect imitation. The authors would like to implement and evaluate several different ways of computing sizes and reach some conclusions about which strategies are most efficient.

While the greedy algorithm is allowed to search for the fastest path to a counterexample, there may well be more efficient algorithms still. There is still a bias towards finding earlier paths. It would be interesting to see how a random walk would perform.

Currently, the user must explicitly request parallel QuickCheck by using
\code{quickCheckPar} instead of \code{quickCheck}. This choice was made because
properties involving I/O can not always be safely executed in parallel. It would
be possible to instead have QuickCheck automatically execute tests in parallel
when it is safe to do so. For example, pure properties (not using $ioProperty$)
can always be parallelized. Properties doing I/O could be marked as thread-safe using a special combinator.

We are also working together with the QuickCheck maintainers towards merging this line of work into mainline QuickCheck.

\bibliographystyle{ACM-Reference-Format}
\bibliography{paper}

\appendix

\section{The effect of Chatty}
\label{ap:chatty}

The \code{chatty} flag in QuickCheck controls whether QuickCheck should continuously print what it is doing or not. While printing is helpful in assessing the current progress, it can be a bottleneck when it comes to performance.

QuickCheck prints the current progress before every test. If you run just a few tests every second this is of no concern, but if your property is a very fast one it has a huge effect on performance. Running 10000 tests per second means that you will print 10000 times per second, which is significantly more than a human eye can observe.

QuickerCheck takes a different approach to printing. Since the new run-time is multi-threaded anyway, QuickerCheck will spawn a separate worker thread whose sole purpose is to periodically print the progress to the terminal. The duration of the period can be configured, and the default is 200 milliseconds.

\begin{figure}[b]
    \centering
    \includegraphics[scale=0.5]{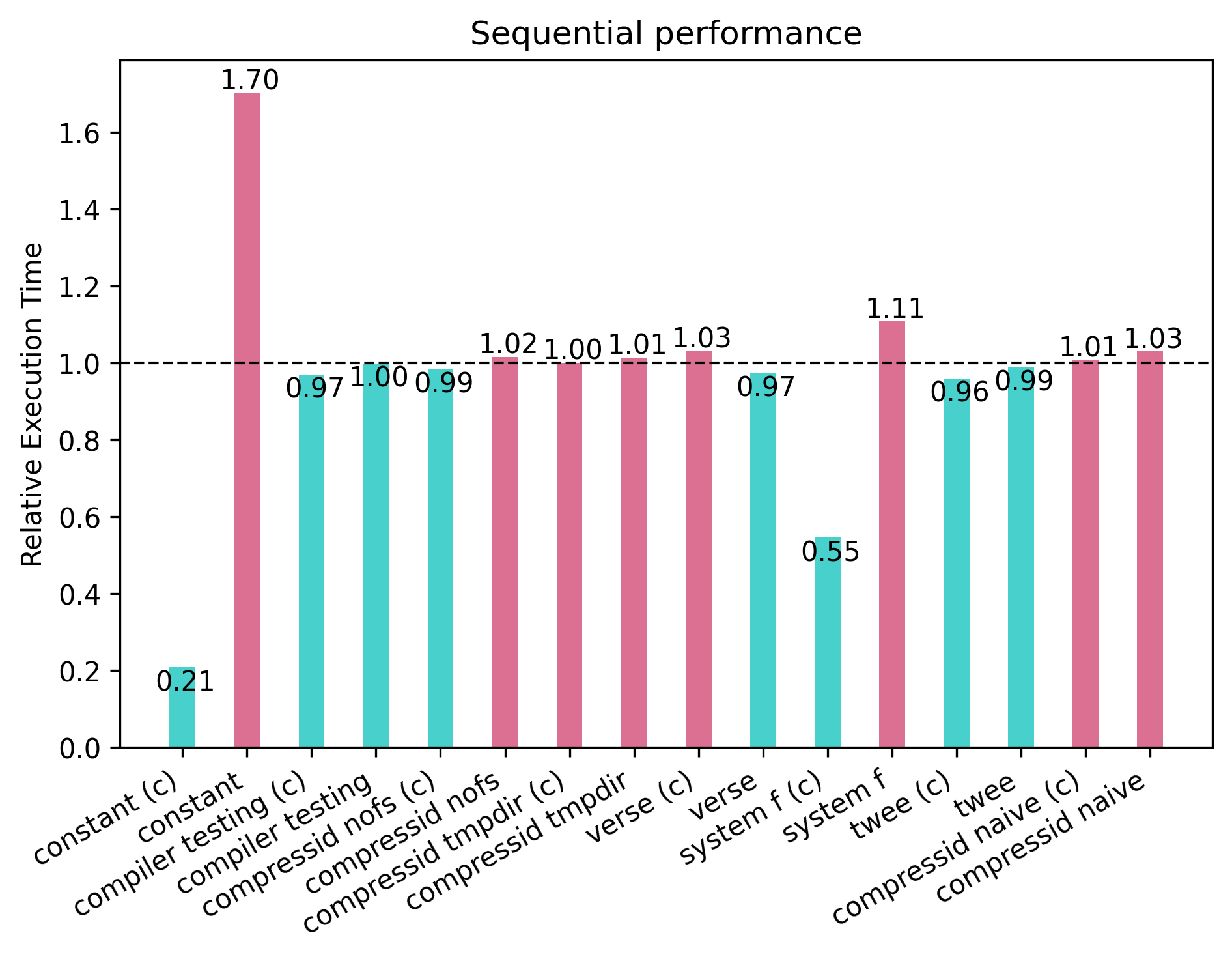}
    \caption{The sequential performance of QuickerCheck relative to QuickCheck, evaluated with the \code{chatty} flag turned on and off. The \textit{(c)} suffix indicates that \code{Chatty} was set to \code{True}.}
    \label{fig:seq_with_chatty}
\end{figure}

While the property that now runs 10000 tests in one second would previously have printed 10000 times, QuickerCheck would only have printed 5 times. In figure \ref{fig:seq_with_chatty} it can be observed how much of an effect this has on a sequential workload. The \textit{constant} and \textit{system f} properties run extremely fast, and we observe that with the \code{chatty} flag set to \code{True}, QuickerCheck is significantly faster. The \textit{constant} property finished evaluating in one-fifth of the time that QuickCheck required.

\begin{figure}[b]
    \centering
    \includegraphics[scale=0.5]{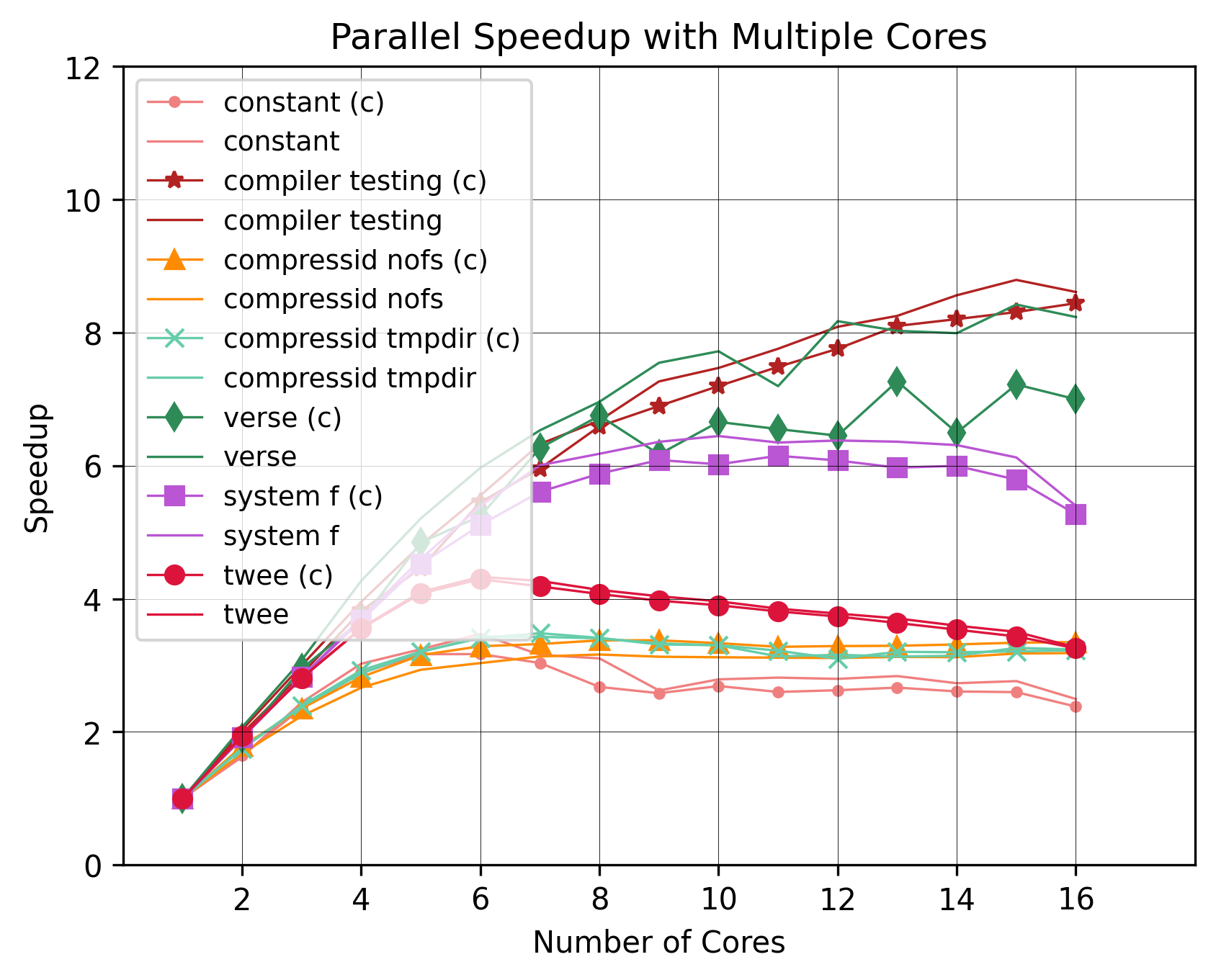}
    \caption{The speedup relative to sequential execution time, when \texttt{chatty} is enabled.}
    \label{fig:speedup_with_chatty}
\end{figure}

As we add cores, it appears that \code{chatty} might make the benchmarks scale slightly worse, but not a lot, as indicated in figure \ref{fig:speedup_with_chatty}.

\end{document}